\documentclass[notitlepage, final, 10pt, prx, aps, twocolumn, floatfix, letter, longbibliography]{revtex4-2}
\pdfoutput=1

\usepackage[utf8]{inputenc}
\usepackage[english]{babel}
\usepackage{lmodern}
\usepackage{xcolor}
\definecolor{navyblue}{rgb}{0.0, 0.0, 0.5}
\usepackage{amsmath, amssymb, amsfonts, bm}
\usepackage{latexsym}
\usepackage{bbm}
\usepackage{mathtools}
\usepackage{placeins}
\usepackage{braket}
\usepackage{booktabs}
\usepackage[final]{changes}
\usepackage[protrusion=true, expansion=true]{microtype}
\usepackage[colorlinks=true, breaklinks=false, linkcolor=navyblue, urlcolor=navyblue, citecolor=navyblue]{hyperref}
\usepackage{graphicx}
\usepackage{float}
\usepackage{siunitx}
\usepackage{csquotes}
\usepackage{cleveref}

\newcommand{\ham}{\hat{H}}

\begin{document}
\title{Beyond hard-core bosons in transmon arrays}

\author{Olli Mansikkamäki}
\author{Sami Laine}
\author{Atte Piltonen}
\author{Matti Silveri}

\affiliation{Nano and Molecular Systems Research Unit, University of Oulu, FI-90014 Oulu, Finland}

\date{\today}

\begin{abstract}
Arrays of transmons have proven to be a viable medium for quantum information science and quantum simulations. Despite their widespread popularity as qubit arrays, there remains yet untapped potential beyond the two-level approximation or, equivalently, the hard-core boson model. With the higher excited levels included, coupled transmons naturally realize the attractive Bose-Hubbard model. The dynamics of the \added{full} model has been difficult to study due to the unfavorable scaling of the dimensionality of the Hilbert space with the system size. In this work, we present a framework for describing the effective unitary dynamics of highly-excited states of coupled transmons based on high-order degenerate perturbation theory. This allows us to describe various collective phenomena -- such as bosons stacked onto a single site behaving as a single particle, edge-localization, and effective longer-range interactions -- in a unified, compact, and accurate manner. \replaced{A further benefit of our approach is that boson stacks can be naturally interpreted as interacting quasiparticles, enabling transmon arrays to be used to explore and study additional lattice models besides the standard Bose-Hubbard one.}{\added{Interacting boson stacks are then modeled as quasiparticles, which enables explorations and studies of behaviors beyond the standard Bose-Hubbard model with transmon arrays.}} While our examples deal with one-dimensional chains of transmons for the sake of clarity, the theory can be readily applied to more general geometries.
\end{abstract}
\maketitle

\section{Introduction}
Superconducting quantum devices have arisen as a practical platform for creating and studying synthetic quantum matter~\cite{carusotto_photonic_2020} through, for example, the realizations of a Mott insulator~\cite{ma_dissipatively_2019} and many-body localization~\cite{Roushan17, chiaro19, Guo21, Gong21, Mi22}, as well as for probing the propagation of single- and many-body quantum information~\cite{ye_propagation_2019, Mi21, braumuller_probing_2021, Karamlou22}. Their remarkable progress has been made possible by universal single-site control, high-accuracy measurements, scalability, and connectivity, all combined with low dissipation and decoherence rates~\cite{kjaergaard_superconducting_2019, Blais21}. There does, however, still remain potential which is largely unutilized. Namely, the Hilbert space beyond the two-level approximation, ignored when the devices are operated as qubits. The total Hilbert-space dimension is a critical resource in quantum computing and quantum simulations~\cite{Blume02}. To increase it, the first strategy is to simply have more precious quantum hardware~\cite{Bernien17, Arute19, Yang20, Zhong20, Gong21_walks}. The second strategy -- the one we are interested in here -- is to utilize larger parts of the Hilbert spaces of the existing physical \replaced{equipment}{resources}.

The most popular superconducting quantum device is the transmon~\cite{Koch07, Paik11}, an anharmonic oscillator whose two lowest states are typically operated as a qubit. In an array, the hopping rate $J$ between the transmons is weak compared to the anharmonicity $U$ of an individual transmon. As a consequence, the quantum dynamics of the system can be predicted with relatively good accuracy using the hard-core boson model~\cite{Yanay20, braumuller_probing_2021}, \emph{provided} the initial state has no higher-level occupancies. In other words, the hard-core boson model is the many-body version of the two-level truncation. Notably, most of the experimental quantum dynamics studies have limited themselves to this case where no transmon is initially excited above the qubit subspace~\cite{Roushan17, Arute19, ye_propagation_2019, braumuller_probing_2021,Gong21, Karamlou22}.

It has been experimentally demonstrated that the higher excited states of a transmon are almost as well controllable and measurable as the ones in the qubit subspace~\cite{Bianchetti10, Peterer15, Elder20, Blok21, Morvan21, Cervera-lierta22}. With the higher levels included, coupled transmons naturally realize the attractive Bose-Hubbard model~\cite{degrandi15, Dalmonte15, Roushan17, mansikkamaki_phases_2021}. As coupled anharmonic oscillators, sets of transmons obey bosonic many-body statistics, yielding vastly larger Hilbert spaces compared to those of the qubit arrays. \replaced{The dynamics of bosonic models can therefore become computationally expensive to study when the system sizes are increased even moderately~\cite{Orell19} compared to the minimal examples presented here.}{Full quantum dynamics of bosonic models is therefore challenging to study \added{numerically}, even with a moderate system size~\cite{Orell19}.} On the other hand, going beyond the hard-core bosons would provide novel experimental and theoretical possibilities to study richer quantum dynamics~\cite{Orell21}. In order to unlock these prospects, we present here a systematic framework for describing effective unitary dynamics of highly-excited states of coupled transmons based on high-order degenerate perturbation theory. \added{Our main focus here is on unitary dynamics, ignoring \added{any} non-unitary effects caused, for example, by dissipation and dephasing processes in transmons. \replaced{To assess the feasibility of this assumption, a brief comparison between the time scales of the examined unitary dynamics and the realistic transmon qubit dissipation and dephasing times $T_1$ and $T_2$ will be given in the final section.}{A time scale comparison between the realistic transmon qubit dissipation and dephasing times $T_1$ and $T_2$ and that of the presented quasiparticle dynamics will discussed in the final section.}}
 
Generally speaking, the Bose-Hubbard model is a well-studied system~\cite{Cazalilla11}. Much of the analysis, however, pertains to optical lattices, in contrast to the superconducting-circuit perspective taken here. Especially the simplest case of a pair of bosons has garnered a wealth of theoretical \cite{stefanak_directional_2011, ahlbrecht_molecular_2012, lahini_quantum_2012, gorlach_topological_2017, stepanenko_interaction-induced_2020} and experimental \cite{folling_direct_2007, winkler_repulsively_2006, preiss_strongly_2015} attention. We concentrate here on exploring the collective dynamics of stacks of multiple bosons, \added{which can be identified as quasiparticles}, and their interactions both with each other and with single bosons. Some many-body effects, such as edge-localization \cite{pinto_edge-localized_2009}, correlated hopping \cite{compagno_noon_2017}, and effective longer-range interactions \cite{bissbort_effective_2012, valiente_three-body_2010}, have also been previously discussed. All of these effects and more, we argue, can be described accurately, compactly, and in a unified manner using perturbation theory appropriate for experimentally relevant parameter values of transmon arrays. Furthermore, we go beyond the earlier studies by focusing on multiple bosons at single lattice sites, corresponding to the dynamics of highly excited transmon states within the arrays. \added{Note that the many-body energy level spectrum of the repulsive Bose-Hubbard model is a mirror image of that of the attractive one, and in both cases the most intense focus has been on the properties of the ground states and the low-lying excited states~\cite{Cazalilla11, mansikkamaki_phases_2021}. In contrast, we resolve here dynamics and structure of states also in the mid-energy spectrum.}

The paper is organized as follows. Section~\ref{sec:intro2} considers the transmon array, the attractive Bose-Hubbard model, and qualitative characteristics thereof within the parameter regime relevant to superconducting quantum devices. In Sec.~\ref{sec:theory}, we introduce the formalism of high-order degenerate perturbation theory, which is then applied in Secs.~\ref{sec:single_stack}--\ref{sec:stacks} to derive effective Hamiltonians and unitary quantum dynamics for higher excited states, that is, for multiple bosons, in 1D transmon arrays. Our focus is on collective effects, such as bosons stacked onto a single site behaving as a single particle, edge-localization, and effective longer-range interactions. In Sec.~\ref{sec:2D}, we discuss \added{in more detail the effects of realistic disorder in transmon arrays} and the generalization \added{of the phenomena} to 2D arrays before concluding in Sec.~\ref{sec:conc}. The main text concentrates on the physical concepts and phenomena, whereas details and derivations are given in Apps.~\ref{app:theory}--\ref{app:numerics}.

\section{Transmon arrays and the attractive Bose-Hubbard model} \label{sec:intro2}
\begin{figure}
    \centering
    \includegraphics[width=\columnwidth]{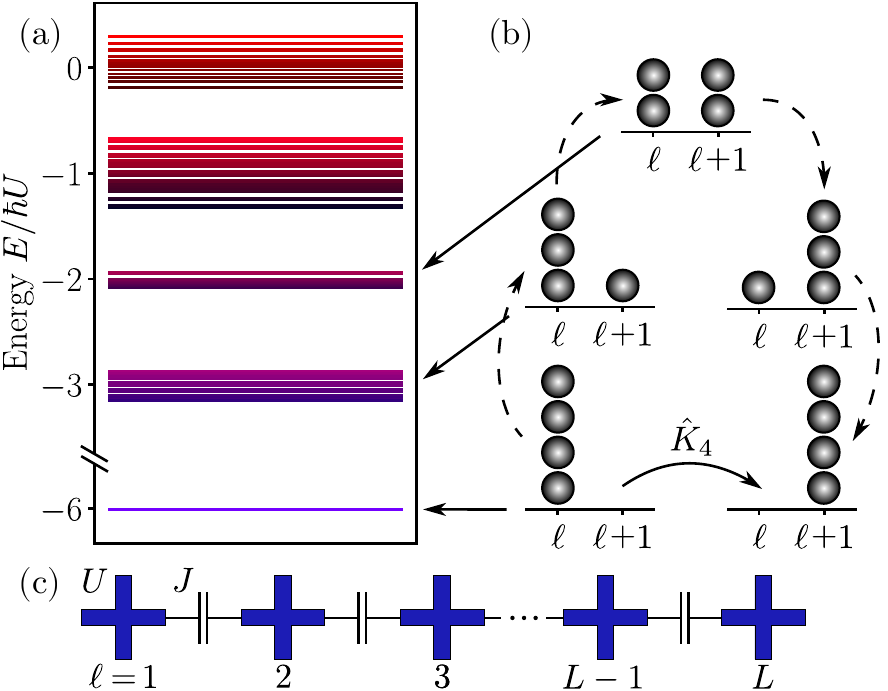}
    \caption{(a)~The energy level spectrum of a transmon chain of length $L = 6$ containing a total of $N = 4$ bosons, as described by the attractive Bose-Hubbard model (\ref{eq:ham}). In this work, we are interested in the many-body dynamics beyond the hard-core bosons, meaning that we consider also the states below the topmost energy band. (b)~A schematic representation of an effective hop of a stack of bosons described by the operator $\hat{K}_4$ [see Eq.\ (\ref{eq:K_n})] within the lowest anharmonicity manifold, together with the virtual hops through the higher manifolds involved in the process. (c)~A schematic of a chain of $L$ capacitively coupled transmons.}
    \label{fig:spectrum}
\end{figure}

An array of superconducting transmon devices [Fig.~\ref{fig:spectrum}(c)] can be described using the disordered Bose-Hubbard model with attractive interactions \cite{degrandi15, Dalmonte15, Roushan17, Orell19, mansikkamaki_phases_2021}, defined by the Hamiltonian
\begin{align}\label{eq:ham}
    \ham & = \ham_J + \ham_U + \ham_\omega \\
    & =\sum_{\langle \ell_1, \ell_2 \rangle} \hbar J^{}_{\ell_1 \ell_2} \hat{a}^\dagger_{\ell_1} \hat{a}^{}_{\ell_2} - \sum_{\ell} \frac{\hbar U_{\ell}}{2} \hat{n}_\ell (\hat{n}_\ell - 1) + \sum_{\ell} \hbar \omega_{\ell} \hat{n}_\ell \notag
\end{align}
when written in the basis of local bosonic annihilation $\hat{a}_{\ell}$, creation $\hat{a}^\dagger_{\ell}$, and occupation number $\hat{n}_{\ell} = \hat{a}^\dagger_{\ell} \hat{a}^{}_{\ell}$ operators. Here, $\ham_J$ allows the excitations to move from one lattice site to a neighboring one, $\ham_U$ takes into account local attractive interactions between the excitations, and $\ham_\omega$ represents the potential energy landscape provided to the excitations by the transmon array. Typically~\cite{Koch07, Paik11, Arute19}, the on-site energy $\omega_\ell / 2 \pi \sim \SI{5}{\giga\hertz}$, the interaction strength $U_\ell / 2 \pi \sim \SI{200}{\mega\hertz}$, and the hopping rate $J_{\ell_1 \ell_2} / 2 \pi \sim \SI{10}{\mega\hertz}$. The parameters of the model therefore satisfy the hierarchy $J_{\ell_1 \ell_2} \ll U_\ell \ll \omega_\ell$. Finally, $\hbar$ is the reduced Planck's constant.

For the majority of this paper, we concern ourselves with a chain of length $L$ of identical transmons with constant nearest-neighbor hopping rate, that is, we set $\omega_\ell = \omega$, $U_\ell = U$, and $J_{\ell_1 \ell_2} = J$. In reality, small unintentional variations in the manufacturing process are inevitable, leading to some disorder in all the parameters.
\replaced{
Since the on-site energy $\omega$ dominates the interaction $U$ and the hopping $J$, \emph{a priori} the most significant disorder term is $\ham_{\delta \omega} / \hbar = \sum_\ell \delta \omega_\ell \hat{n}_\ell$. Here the distribution of the deviations $\delta \omega_\ell = \omega_\ell - \omega$ can be approximated as uniform \cite{Orell19}, with the half-width $D_\omega$ measuring the disorder strength. With flux tuning, however, the on-site energies can be adjusted, and values as low as $D_\omega / 2 \pi \sim \SI{100}{\kilo\hertz}$ are currently achievable~\cite{ma_dissipatively_2019}, leading to $D_\omega \ll J$. It is therefore more likely that disorder in $U$, which is essentially impossible to control after the device is made, is the actual limiting factor. The corresponding term in the Hamiltonian is given by $\ham_{\delta U} / \hbar = - \sum_{\ell} \delta U_\ell \hat{n}_\ell (\hat{n}_\ell - 1) / 2$, and at least in current devices the disorder strength characterizing the deviations $\delta U_\ell = U_\ell - U$ is usually no less than $D_U / 2 \pi \sim \SI{1}{\mega\hertz}$. Nevertheless, with \emph{ad hoc} tuning of the on-site energies $\omega_\ell$, the effective disorder in the interaction strength $U_\ell$ can -- at least in some cases -- be made smaller. We shall discuss this briefly in Sec.\ \ref{sec:2D}, after the analysis of the ideal system.
}
{
Since the on-site energy $\omega$ dominates the interaction $U$ and the hopping $J$, the most significant disorder term is $\ham_D / \hbar = \sum_\ell (\omega_\ell - \omega) \hat{n}_\ell$. The distribution of the deviations $\omega_\ell - \omega$ can be approximated as uniform \cite{Orell19}, with the half-width $D$ measuring the disorder strength. With flux tuning, however, values as low as $D / 2 \pi \sim \SI{100}{\kilo\hertz}$ are currently achievable~\cite{ma_dissipatively_2019}, and thus $D \ll J$. It is therefore more likely that disorder in $U$, which is essentially impossible to control after the device is made, is actually the limiting factor. Nevertheless, the effects caused by disorder in either $\omega$ or $U$ are similar, and so we simply refer to the disorder $D$. We shall briefly discuss the qualitative effects of the disorder after the analysis of the ideal system.
}

The model~\eqref{eq:ham} conserves the total number of bosons $\hat{N} = \sum_{\ell} \hat{n}_\ell$, and thus, in the absence of dissipation, we can separately study the eigenspaces $\hat{N} = N$. Since $\ham_\omega \propto \hat{N}$, it simply shifts all the energies by the same amount without affecting the states, and can therefore be omitted.

Under the assumption $J \ll U$, the total anharmonicity $\hat{A} = -\sum_{\ell} \hat{n}_\ell (\hat{n}_\ell - 1)/2$ is approximately conserved. This key observation helps us simplify the analysis considerably. To see why, let us consider the energy spectrum of the system, depicted in Fig.\ \ref{fig:spectrum}(a). If the hopping frequency $J$ vanished altogether, we would have $\ham = \hbar U \hat{A}$, and the spectrum would consist of a small number of highly degenerate lines determined by the possible values $A$ of the anharmonicity operator $\hat{A}$. Each such anharmonicity manifold is spanned by the Fock states $\ket{n_1, n_2, \ldots, n_L}$, with $n_\ell$ bosons at site $\ell$, satisfying the conditions $\sum_{\ell} n_\ell = N$ and $\sum_{\ell} n_\ell (n_\ell - 1) = -2 A$. The lowest anharmonicity $A = -N(N - 1)/2$, and thus also the lowest energy, is achieved if all the bosons sit at a single site. We denote these states with the shorthand notation $\ket{N_\ell}$, omitting all the zero occupations. Second lowest in anharmonicity are the states $\ket{(N - 1)_\ell, 1_m}$ where all but one of the bosons are located at the same site. The greatest anharmonicity, $A = 0$, belongs to the states where no site is occupied by more than one boson.

Turning on $0 < J \ll U$, the main effect of the hopping $\ham_J$ is to couple together the states \emph{within} each anharmonicity manifold. This lifts the degeneracies, leading to a spectrum with discrete, well separated bands around the energies $\hbar U A$, as shown in Fig.\ \ref{fig:spectrum}(a). To leading order, the states within each such band still belong to the corresponding anharmonicity manifold while different bands remain uncoupled. Importantly, unlike in the absence of hopping when a state with definite $A$ does not evolve in time apart from a trivial phase factor, there can now be nontrivial dynamics within each anharmonicity manifold. Since $\ham_J$ can only move one boson to an adjacent site, it is not necessarily able to couple states with equal $A$ directly. For example, to couple the states $\ket{N_\ell}$ within the the lowest anharmonicity manifold requires at least $N$ one-boson hops. This means that the coupling has to be indirect, mediated by the states in the other anharmonicity manifolds as depicted by the dashed arrows in Fig.~\ref{fig:spectrum}(b). Each jump needed decreases the magnitude of the coupling strength by a factor of $J/U$. The nearest-neighbor coupling strength between the states $\ket{N_\ell}$ is therefore~$\sim U (J/U)^N$. The higher-order coupling, in turn, leads to dynamics which is slow even compared to $J$, and hence it is easy to overlook curious phenomena if not careful.

When considering dynamics, the relevant anharmonicity manifolds are picked by the initial state. A lot of attention has been paid to the highest manifold with no multi-boson occupancies at any site. This situation can be effectively described with the so-called hard-core boson model~\cite{Yanay20, braumuller_probing_2021}. But by choosing the initial state appropriately, one can equally well study any of the other manifolds, each with a different effective model.

In this article, we derive general expressions for the effective Hamiltonians describing the dynamics induced by the weak hopping within each anharmonicity manifold. These are especially convenient when combined with numerical analysis since the reduced Hilbert space size allows simulation of bigger systems when compared to the full Hamiltonian. In addition, a lot of intuition regarding the qualitative physics of the system can be gained by examining the effective models analytically. We apply the effective theory to study the lower end of the spectrum, where one can observe various interesting many-body effects. For the sake of clarity, we focus on dynamics starting from initial states where only one or two sites are occupied by some number of excitations. It should be noted, however, that the same methods, and indeed some of the results, can be directly applied to cases with three or more occupied sites.

\section{High-order degenerate perturbation theory} \label{sec:theory}
Since the Hamiltonian $\ham$ of Eq.\ (\ref{eq:ham}) is independent of time, any initial state $\ket{\Psi_0}$ is propagated by the exponential $e^{-i \ham t / \hbar}$. Denoting the eigenvalues and eigenstates of $\ham$ by $E$ and $\ket{E}$, respectively, we can write the state of the system at time $t$ as $\ket{\Psi(t)} = \sum_E e^{-i E t / \hbar} \braket{E | \Psi_0} \ket{E}$. This is an exact expression but requires solving the full eigenvalue problem. To proceed, we make use of the assumption $J \ll U$ and expand both the energies and the states in powers of $J/U$ as \replaced[comment=Testi]{
\begin{align}
E &= \sum_{n = 0}^{\infty} E^{(n)},&  \ket{E} &= \sum_{n = 0}^{\infty} \ket{E^{(n)}},
\end{align}
}{$E = \sum_{n = 0}^{\infty} E^{(n)}$ and $\ket{E} = \sum_{n = 0}^{\infty} \ket{E^{(n)}}$,} respectively, with $E^{(n)}, \ket{E^{(n)}} \sim (J/U)^n$. Keeping only the leading-order term in \emph{amplitude}, we obtain
\begin{equation}\label{eq:psi(t)_approx}
    \ket{\Psi(t)} = e^{-i \ham_{\mathrm{eff}} t / \hbar}  \ket{\Psi_0} + \mathcal{O}\left(J / U \right),
\end{equation}
where the effective Hamiltonian
\begin{equation}\label{eq:eff_ham}
    \ham_{\mathrm{eff}} = \sum_{E} E \ket{E^{(0)}} \bra{E^{(0)}}
\end{equation}
generates the dominant part of the dynamics. Note that at this point, we still retain the full energy inside $\ham_{\mathrm{eff}}$. This is crucial since it affects the \emph{phase} of the state rather than the magnitude. Depending on the initial state, the system may possess interesting physics at slower time scales, and thus truncating the energies too early completely wipes out these phenomena from the analysis. The higher-order states, on the other hand, only contribute to $\ket{\Psi(t)}$ with terms which are small in magnitude, and can thus be safely omitted.

The task we face, then, is to calculate the zeroth-order eigenstates of the Hamiltonian (\ref{eq:ham}) together with the corresponding energies to a sufficiently high order so as not to miss anything important while simultaneously keeping in mind experimental limitations. As we shall see, these two problems are closely intertwined.

The unperturbed Hamiltonian is given by the interaction term $\ham_U \propto \hat{A}$, and so we know that each state $\ket{E^{(0)}}$ has a definite anharmonicity, allowing us to concentrate on one anharmonicity manifold at a time. Note that only the states with a component along the initial state are important. Starting from a Fock state, for example, requires just a single anharmonicity manifold to be taken into account.

To proceed, we employ the standard degenerate perturbation theory \cite{mansikkamaki_phases_2021, sakurai2017}, see App.\ \ref{app:theory} for details. To this end, let $\mathcal{A}$ be the anharmonicity manifold we are interested in and let $\mathcal{A}^c$ be its complement. Furthermore, we define the projectors $\hat{P}_\mathcal{A}$ and $\hat{Q}_\mathcal{A} = \hat{I} - \hat{P}_\mathcal{A}$ to the spaces $\mathcal{A}$ and $\mathcal{A}^c$, respectively, with $\hat{I}$ denoting the identity operator. Projecting the time-independent Schrödinger equation into $\mathcal{A}$ then leads to the equation
\begin{equation}\label{eq:proj_schrödinger}
    \ham_\mathcal{A}(E) \ket{E_\mathcal{A}} = E \ket{E_\mathcal{A}},
\end{equation}
with $\ket{E_\mathcal{A}} = \hat{P}_\mathcal{A} \ket{E}$. The projected Hamiltonian $\ham_\mathcal{A}(E)$ can be written as
\begin{equation}\label{eq:proj_ham}
    \ham_\mathcal{A}(E) = \hbar U A + \sum_{m = 1}^{\infty} \hat{K}_m(E),
\end{equation}
where $A$ is the anharmonicity of $\mathcal{A}$ and
\begin{align}
    \hat{K}_m(E) &= \hat{P}_\mathcal{A} \ham_J [\hat{W}(E) \ham_J ]^{m - 1} \hat{P}_\mathcal{A}, \label{eq:K_n} \\
    \hat{W}(E) &= [\hat{Q}_\mathcal{A} (E - \ham_U ) \hat{Q}_\mathcal{A} ]^{-1}. \label{eq:W}
\end{align}
The operator $\hat{K}_m$ can be interpreted as a weighted $m$th-order hopping Hamiltonian since operating with $\hat{K}_m$ on a Fock state from $\mathcal{A}$ produces a linear combination of all the possible Fock states within $\mathcal{A}$ which are exactly $m$ single-boson hops away from the original state. Each $m$-hop sequence or \emph{trajectory} connecting two states contributes to the corresponding weight in the linear combination. More weight is given to trajectories for which the intermediate states are close in energy to the actual value $E$ (measured using the unperturbed Hamiltonian). Moreover, and perhaps more importantly, all the intermediate states on a trajectory need to lie in $\mathcal{A}^c$ due to the presence of the projectors $\hat{Q}_\mathcal{A}$ in Eq.\ (\ref{eq:W}), otherwise the weight is zero. This allows for a nice graphical way to understand the most important terms in Eq.\ (\ref{eq:proj_schrödinger}) via virtual hopping processes, see Fig.~\ref{fig:spectrum}(b). A similar diagrammatic approach has been used to calculate various ground-state expectation values in the repulsive Bose-Hubbard model \cite{eckardt2009, teichmann2009a, teichmann2009b, heil2012, hinrichs2013, wang2018, kubler2019, sanders2019, wang2020} and other lattice models \cite{eckardt2009, heil2012, kalinowski2012}.

Equation (\ref{eq:proj_schrödinger}) is still exact. It can be used to calculate the full energies $E$ of such eigenstates that belong to $\mathcal{A}$ in the limit of vanishing $J$, and at the same time, it gives the projections of the states inside $\mathcal{A}$. If necessary, one can then use these to obtain also the components of the states in $\mathcal{A}^c$. We are here only interested in the zeroth-order states which always lie in $\mathcal{A}$, and thus Eq.\ (\ref{eq:proj_schrödinger}) is sufficient. Note that the projected Hamiltonian depends on the energy, and thus the problem is nonlinear. The dimensionality of the Hilbert space is, however, reduced from the original value of $\binom{N + L - 1}{N}$ to the dimensionality of $\mathcal{A}$.

If we now let $\ham_\mathcal{A}^{(n)} \sim (J/U)^n$ denote the part of the projected Hamiltonian that is obtained by expanding $\ham_\mathcal{A}(E)$ in energy and keeping all the terms which are of $n$th order in $J / U$, Eq.\ (\ref{eq:proj_schrödinger}) implies that the zeroth-order eigenstates can be solved iteratively from
\begin{equation}\label{eq:schrödinger_n}
    \ham_\mathcal{A}^{(n)} \ket{E^{(0)}} = E^{(n)} \ket{E^{(0)}},
\end{equation}
where $n = 1, 2, \ldots$ denotes the order. First, we solve the eigenvalue problem of $\ham_\mathcal{A}^{(1)}$. If there are eigenvalues which are nondegenerate, the corresponding states are proper eigenstates and they need no further treatment. If there are still degeneracies left, we proceed to second order. There, we can solve the eigenvalues and eigenstates of $\ham_\mathcal{A}^{(2)}$ separately in all the different degenerate eigenspaces of $\ham_\mathcal{A}^{(1)}$, again turning a single eigenvalue problem into multiple smaller ones. This procedure is then continued until all the degeneracy is lifted or we reach such a high order that the time scale required to observe the resulting dynamics is too slow to be relevant. Note that the process yields the energy of each state up to the order where the degeneracy with the rest of the states is broken. These are the natural cut-off points in the sense that moving to higher orders reveals no fundamentally new physics in the system. \deleted{For more details of the high-order degenerate perturbation theory, see App.~\ref{app:theory}.}

In the following, we consider a few examples both in order to demonstrate the power of the effective Hamiltonian method presented here and to showcase some interesting phenomena found in transmon arrays beyond the qubit approximation.

\section{Collective motion of a boson stack}\label{sec:single_stack}
\begin{figure}
    \centering
    \includegraphics[width=\columnwidth]{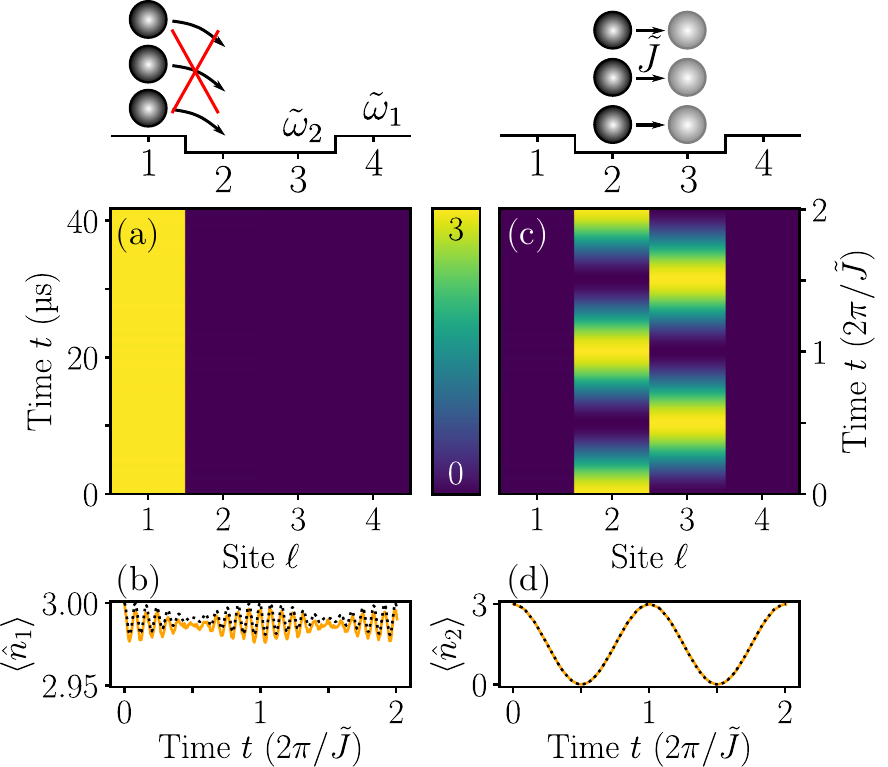}
    \caption{Edge localization and slow collective hopping of a boson stack. (a, c) The local occupations $\langle \hat{n}_\ell \rangle$ as a function of the site index $\ell$ and time $t$. (b, d) The occupation of the site $\ell = 1$ (b) and the site $\ell = 2$ (d) as a function of time $t$, simulated using the full Hamiltonian of Eq.\ \eqref{eq:ham} (solid yellow) and the effective Hamiltonian of Eq.\ \eqref{eq:effective_hamiltonian} (dotted black). The initial state is $\ket{3_1} = \ket{3000}$ in (a--b) and $\ket{3_2} = \ket{0300}$ in (c--d). The simulation parameters in this and all the following figures are $J/2\pi = \SI{10}{\mega\hertz}$ and $U/2\pi = \SI{250}{\mega\hertz}$. Note the two time axes in (a) and (c), one in SI units (left axis) and the other in the natural units $2 \pi / \tilde{J}$ of the effective hopping rate (right axis).}
    \label{fig:single_stack}
\end{figure}

Let us first consider dynamics starting from an initial state where all the $N$ bosons in the system are stacked onto the same site $\ell_0$, that is, $\ket{\Psi_0} = \ket{N_{\ell_0}}$. In this case, the only relevant anharmonicity manifold $\mathcal{A}$ is the lowest one, spanned by the states $\ket{N_\ell}$ with $\ell = 1, \ldots, L$. Note that the dimensionality of $\mathcal{A}$ is independent of $N$ and grows linearly in $L$, as opposed to the exponential growth of the full Hilbert space. The analysis here is closely related to the calculation of the ground state of the model within the W phase \cite{mansikkamaki_phases_2021}.

For the sake of convenience, let us define new creation and annihilation operators $\hat{\alpha}_\ell^\dagger$ and $\hat{\alpha}_\ell$ that create and destroy the whole stack at site $\ell$, respectively. That is, $\hat{\alpha}_\ell^\dagger \ket{0} = \ket{N_\ell}$, where $\ket{0}$ denotes the vacuum state with no excitations. The corresponding number operator, counting the number of $N$-particle stacks instead of the number of bosons, is defined as $\hat{\nu}_\ell = \hat{\alpha}_\ell^\dagger \hat{\alpha}_\ell$.

Following then the procedure set out in the previous section -- see App.~\ref{app:single_stack} for details -- we find that the effective Hamiltonian consists of only two relevant parts. First of all, the stack is able to move around just like an individual boson due to the presence of the nearest-neighbor hopping Hamiltonian
\begin{equation}\label{eq:effective_hopping}
\ham_{\tilde{J}} / \hbar = \tilde{J} \sum_{\ell = 1}^{L - 1} (\hat{\alpha}_{\ell + 1}^\dagger \hat{\alpha}^{}_\ell + \hat{\alpha}_\ell^\dagger \hat{\alpha}^{}_{\ell + 1} ),
\end{equation}
where the effective hopping frequency is given by
\begin{equation}\label{eq:tilde_J}
    \tilde{J} = \replaced{(-1)^{N - 1} \frac{N}{(N - 1)!} \left( \frac{J}{U} \right)^{N - 1} J}{(-1)^{N - 1} \frac{N}{(N - 1)!} \left( \frac{J}{U} \right)^N U}.
\end{equation}
We can therefore think of the stack as a single \emph{quasiparticle} whose effective mass increases exponentially as a function of $N$. Unlike the bare bosons, however, the quasiparticle sees the transmon chain as inhomogeneous due to \emph{boundary effects}. More precisely, the effective Hamiltonian contains the on-site term
\begin{equation}\label{eq:boundary_term}
\ham_{\tilde{\omega}} / \hbar = \sum_{\ell = 1}^{L} \tilde{\omega}_{|\ell - \ell_b|} \hat{\nu}_\ell,
\end{equation}
where $|\ell - \ell_b|$ denotes the shortest distance from site $\ell$ to the boundary (for example, $|1 - \ell_b| = |L - \ell_b| = 1$) and, to leading order, the effective on-site frequencies $\tilde{\omega}_\ell$ satisfy
\begin{equation}\label{eq:tilde_omega_diff}
    \tilde{\omega}_\ell - \tilde{\omega}_{\ell + 1} = \replaced{\frac{N}{(N - 1)^{2\ell - 1}} \left(\frac{J}{U}\right)^{2 \ell - 1} J}{\frac{N}{(N - 1)^{2\ell - 1}} \left(\frac{J}{U}\right)^{2 \ell} U}.
\end{equation}
Note that the boundary effects and the effective hopping behave differently as a function of $N$.

Using the effective Hamiltonian
\begin{equation}\label{eq:effective_hamiltonian}
    \ham_{\mathrm{eff}} = \ham_{\tilde{J}} + \ham_{\tilde{\omega}},
\end{equation}
it is now easy to solve qualitatively the dynamics of the system in the lowest anharmonicity manifold spanned by the states~$\ket{N_\ell}$. The quasiparticle can move within a region where it has enough kinetic energy to overcome the local potential energy difference between the neighboring sites. Thus, if the initial site $\ell_0$ is within the distance $\lceil N/2 \rceil - 1$ from a boundary, it is frozen in place. Otherwise, it moves among the $L - 2 (\lceil N/2 \rceil - 1)$ middlemost sites. This motion is practically free since the deviations in $\tilde{\omega}_\ell$ decrease exponentially when moving away from the boundary. Numerical simulations using the full Hamiltonian (\ref{eq:ham}) confirm this behavior, see Fig.\ \ref{fig:single_stack}. By setting $N = 2$, we recover the well-known doublon dynamics \cite{stefanak_directional_2011, ahlbrecht_molecular_2012, lahini_quantum_2012, gorlach_topological_2017, stepanenko_interaction-induced_2020, folling_direct_2007, winkler_repulsively_2006, preiss_strongly_2015}. Note that these do not possess bound states since the depth of the potential well at the end of the chain is of the same order as the effective hopping frequency. Still, the edges affect the motion to some degree. Edge-localization, too, is a familiar phenomenon \cite{pinto_edge-localized_2009}. We could not, however, find any explicit mention of larger stacks being able to localize also further away from the edges. The dynamics for $N = 3$ has been discussed, for example, in Refs.\ \cite{compagno_noon_2017, valiente_three-body_2010}. 

It is perhaps worth mentioning that in principle, there is also a term in the effective Hamiltonian which can move the quasiparticle trapped at site $\ell_0$ near one end of the chain to the corresponding site $L - \ell_0 + 1$ near the other end, but the strength of this coupling is of order $N |L - 2 \ell_0 + 1|$. Even at its fastest, such a phenomenon would require time intervals of $\sim \SI{1}{\second}$ to be observable, and is therefore too slow to be of practical importance, at least for now. \deleted{As a final note, possible disorder can simply be included in the coefficients $\tilde{\omega}_\ell$. Thus, if $D$ is larger than $\tilde{J}$, the dynamics is completely frozen. This applies also more generally: disorder freezes all the motion slower than $\sim 2\pi/D$.}

The origins of both $\ham_{\tilde{J}}$ and $\ham_{\tilde{\omega}}$ can be understood intuitively by considering the virtual hopping processes discussed in the context of Eq.\ (\ref{eq:proj_ham}), see also Fig.~\ref{fig:spectrum}(b). The simplest trajectories connecting two states in $\mathcal{A}$ are the ones where we simply transfer all the $N$ bosons at some site, one by one, to an adjacent site. These give rise to $\ham_{\tilde{J}}$. There are of course more complicated trajectories establishing similar nearest-neighbor couplings, but these require at least $N + 2$ single-boson hops, leading to much weaker -- and consequently insignificant -- coupling strengths. Likewise, the longer-range hopping processes directly connecting sites farther apart are too weak to have any practical effect.

In addition to figuring out how different states are connected to one another, we also need to analyze trajectories connecting each state to itself, leading to $\ham_{\tilde{\omega}}$. If instead of an open chain we had a closed ring, then every site would be on an identical footing. Whatever the actual self-coupling strengths were, they would always be equal, and thus the on-site term would simply reduce to some irrelevant constant. \replaced{In a chain, however, this is not the case since the boundaries break the symmetry of the system by rendering those hopping processes where a boson leaves the chain impossible. To see this more concretely, let us consider the second-order corrections to the diagonal energies. These stem from all such trajectories where we first take one boson from the stack to an adjacent site and then back. In the bulk of the chain, each site has two neighbors, left and right. The end sites, on the other hand, have only one neighbor each. This difference in the number of available adjacent lattice sites leads to higher energy at the edges. More generally, consider the states $\ket{N_\ell}$ and $\ket{N_{\ell + 1}}$ at order $2\ell$, with $\ell < \lceil L/2 \rceil$.}{In a chain, however, this is not the case. To see this, consider the states $\ket{N_\ell}$ and $\ket{N_{\ell + 1}}$ at order $2\ell$, with $\ell < \lceil L/2 \rceil$.} There is now a single trajectory distinguishing these two, impossible for the former state due to the presence of the left \replaced{edge}{boundary}: the one where a lone boson first travels $\ell$ sites to the left and then comes back. All the other ones are identical. This leads to the energy at site $\ell$ being higher than at site $\ell + 1$. Needless to say, the same conclusions can also be drawn for adjacent sites near the other \replaced{end the chain}{boundary}.

\section{Interplay between a boson stack and a single boson} \label{sec:stack_et_boson}
We saw above that a stack of $N$ bosons can be treated as a single massive quasiparticle. How do these quasiparticles interact with individual bosons? To answer this, let us consider the initial state $\ket{\Psi_0} = \ket{N_{\ell_{N 0}}, 1_{\ell_{1 0}}}$. The anharmonicity manifold $\mathcal{A}$ is now spanned by the states $\ket{N_{\ell_N}, 1_{\ell_1}}$, where $\ell_N, \ell_1 = 1, \ldots, L$ and $\ell_1 \neq \ell_N$. This means that it is indeed well-founded to refer to the quasiparticle and the boson throughout the time evolution.

The effective Hamiltonian can be split into two parts. To start with, there are the single-particle Hamiltonians $\ham_J$ and $\ham_{\tilde{\omega}} + \ham_{\tilde{J}}$ generating the free motion of the boson and the quasiparticle, respectively. In addition, there are interactions between the particles strongly modifying the dynamics. Due to the very structure of $\mathcal{A}$, the boson and the quasiparticle cannot occupy a same site, and so there is a hard-core repulsion between the two. This can be modelled with the effective on-site interaction
\begin{equation}
    \ham_{\tilde{U}} / \hbar = \frac{\tilde{U}}{2} \sum_{\ell = 1}^{L} \hat{n}_\ell \hat{\nu}_\ell,
\end{equation}
where $\tilde{U} \to \infty$. Similar to the always-on $Z Z$ interaction of the qubits~\cite{Dicarlo09, Omalley15}, there is also a nearest-neighbor interaction 
\begin{equation}\label{eq:nearest_neighbor_interaction}
    \ham_V / \hbar = V \sum_{\ell = 1}^{L - 1} (\hat{n}_{\ell + 1} \hat{\nu}_\ell + \hat{\nu}_{\ell + 1} \hat{n}_\ell)
\end{equation}
between the particles, with the interaction strength
\begin{equation}
    V = -\left(\frac{2 N}{N - 2} - \frac{N + 1}{N} - \frac{N}{N - 1} \right) \left(\frac{J}{U} \right) J.
\end{equation}
In the case of $N = 2$, we need to drop out the term with the vanishing denominator. Finally, we have two interesting coupling terms. The tunneling Hamiltonian 
\begin{equation}\label{eq:tunneling_term}
    \ham_T / \hbar = T \sum_{\ell = 2}^{L - 1} (\hat{a}_{\ell + 1}^\dagger \hat{\nu}^{}_\ell \hat{a}^{}_{\ell - 1} + \hat{a}_{\ell - 1}^\dagger \hat{\nu}^{}_\ell \hat{a}^{}_{\ell + 1} )
\end{equation}
makes it possible for the boson to hop over the quasiparticle. The tunneling rate is given by
\begin{equation}
    T = - \frac{1}{N(N - 1)} \left(\frac{J}{U} \right) J.
\end{equation}
The exchange Hamiltonian
\begin{equation}\label{eq:exchange_stack+boson}
    \ham_\Xi / \hbar = \Xi \sum_{\ell = 1}^{L - 1} (\hat{a}_{\ell + 1}^\dagger \hat{\alpha}_\ell^\dagger \hat{a}^{}_\ell \hat{\alpha}^{}_{\ell + 1} + \hat{a}_\ell^\dagger \hat{\alpha}_{\ell + 1}^\dagger \hat{a}^{}_{\ell + 1} \hat{\alpha}^{}_\ell ),
\end{equation}
on the other hand, allows neighboring particles to swap sites. Here, the exchange rate is 
\begin{equation}
    \Xi = (-1)^N \frac{N (N - 1)}{(N - 2)!} \left(\frac{J}{U} \right)^{N - 2} J.
\end{equation}
There is one special case, namely $\ell_{N 0} = L/2 + 1$ and $N > 3$, when the effective Hamiltonian contains additional terms, but even then the qualitative physics can be understood well using the terms above as we shall see below. More details can be found in App.~\ref{app:stack+boson}.

\begin{figure}
    \centering
    \includegraphics[width=\columnwidth]{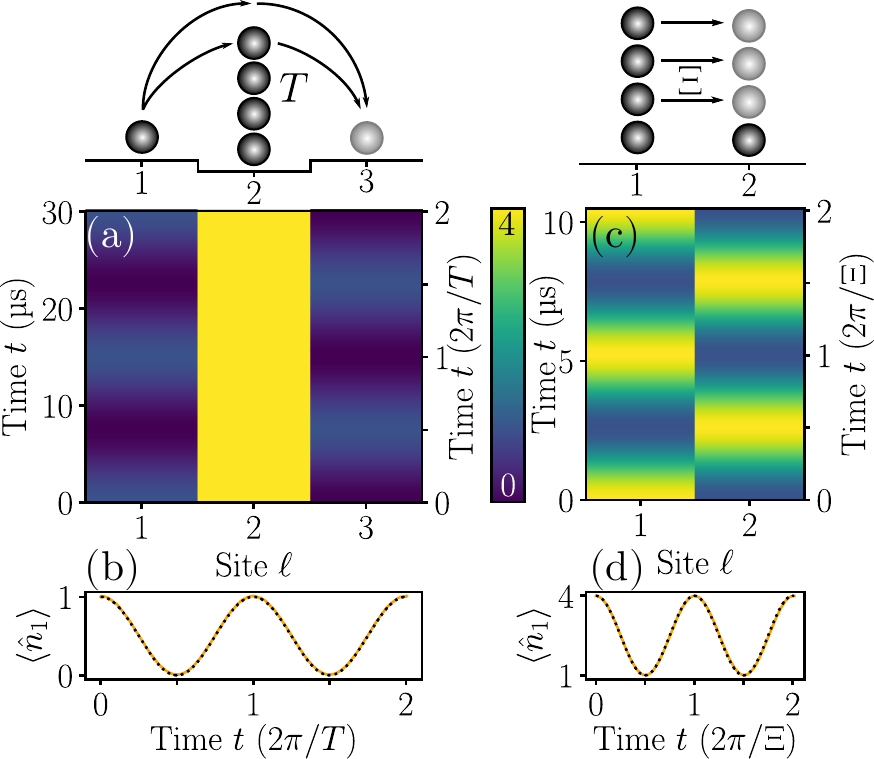}
    \caption{Single-boson tunneling and boson-quasiparticle exchange. (a, c) The local occupations $\langle \hat{n}_\ell \rangle$ as a function of the site index $\ell$ and time $t$. (b, d) The occupation of the site $\ell = 1$ as a function of time $t$, simulated using the full Hamiltonian of Eq.\ \eqref{eq:ham} (solid yellow) and the effective Hamiltonian of either Eq.\ \eqref{eq:tunneling_term} (b) or Eq.\ \eqref{eq:exchange_stack+boson} (d) (dotted black). The initial state is $\ket{140}$ in (a--b) and $\ket{41}$ in (c--d). Note the different time scales involved in the two processes, one determined by the tunneling rate $T$ and the other by the exchange rate $\Xi$.}
    \label{fig:tunneling_and_exchange}
\end{figure}

The dynamics of the system now depends quite strongly on the initial state. Let us first assume that $N > 3$. If the stack is initially at any site $\ell_{N 0} \neq L/2 + 1$, the dynamics includes two time scales. On the time scale of $2\pi / J$, the boson travels freely between the boundary and the quasiparticle, reflecting elastically on impact. On the time scale of $2\pi / T$, the boson can additionally tunnel from one side of the quasiparticle to the other. The actual strength of this tunneling is determined not only by $T$ but also by the initial state, for the energy levels of the boson need to match at least partially on both sides of the effective double well. In the symmetric case of $\ell_{N 0} = (L + 1) / 2$, the levels perfectly coincide and thus the boson can change sides completely. In every other case, mixing of the states between the two wells is much weaker and can, to a decent approximation, be neglected altogether. Figure \ref{fig:tunneling_and_exchange}(a) shows that the analytically predicted tunneling rate matches with the numerical simulations. The exchange plays no role here. Note that the motion of the quasiparticle is completely blocked. This can be understood to large extent with a simple physical argument. Initially, the quasiparticle is delocalized and symmetric in the reciprocal space \cite{mansikkamaki_phases_2021}, meaning that it essentially has zero momentum. In order for the momentum to be conserved, the quasiparticle needs to be able to move in both directions simultaneously. Due to one of its neighboring sites being effectively always occupied by the boson on the time scale of $2\pi / \tilde{J}$, the quasiparticle has to stay still.

The sole exception to the above occurs when $\ell_{N 0} = L/2 + 1$. In this case, the second-order tunneling process is completely prevented by energy conservation. The boson can still swap sides, but this needs to be now accompanied by a simultaneous movement of the quasiparticle, brought about by the exchange term. There are again two time scales involved in the dynamics. As above, the boson bounces back and forth between the quasiparticle and an edge on the time scale of $2\pi / J$. In addition, on the time scale of $2\pi / \Xi$, the quasiparticle moves between the two sites $L/2 + 1$ and $L/2$, exchanging sides with the boson. This is confirmed by numerical simulations, see Fig.\ \ref{fig:tunneling_and_exchange}(b).

If $N = 3$, both tunneling and exchange -- being of the same order -- contribute to the dynamics, making the simple qualitative picture more fuzzy. If $N = 2$, the exchange process is of first order and thus occurs at the same time scale as the motion of the boson. In this case, the quasiparticle can also move longer distances due to the second-order effective hopping $\ham_{\tilde{J}}$, leading again to more involved dynamics.

All the terms above emerge naturally when we think in terms of the virtual hopping processes. The noninteracting terms $\ham_J$ and $\ham_{\tilde{\omega}} + \ham_{\tilde{J}}$ are clear from the earlier discussion. As already mentioned, $\ham_{\tilde{U}}$ stems from the structure of $\mathcal{A}$, for there are no states where all the bosons sit at the same site. The off-site interaction $\ham_V$ has similar origins as the boundary term $\ham_{\tilde{\omega}}$ (see also the discussion in Sec.\ \ref{sec:stacks}): if the particles are at adjacent sites, there are two second-order trajectories coupling the state to itself having an intermediate state where all the $N + 1$ bosons are at the same site, unlike in the case of more distant particles.

To couple states where the boson is on the opposite sides of the quasiparticle, we need at least two virtual hops. The leading-order term, coupling together the states $\ket{N_\ell, 1_{\ell - 1}}$ and $\ket{N_\ell, 1_{\ell + 1}}$, consists of two trajectories. In one of these, we take the boson at site $\ell - 1$, move it to site $\ell$ with all the other bosons, and then move one of these to site $\ell + 1$. In the other, we move one boson from site $\ell$ to site $\ell + 1$, and then move the single boson from site $\ell - 1$ to site $\ell$. These are the only possibilities with two hops, and lead to the tunneling Hamiltonian $\ham_T$. Finally, a completely new coupling, bringing about the exchange term $\ham_\Xi$, can be produced using a trajectory of length $N - 1$. Starting from the state $\ket{N_{\ell}, 1_{\ell \pm 1}}$, we move all but one of the bosons from site $\ell$ to site $\ell \pm 1$, establishing a coupling with the state $\ket{N_{\ell \pm 1}, 1_\ell}$.

\section{Interacting boson stacks} \label{sec:stacks}
As a final example, let us consider the dynamics of two quasiparticles of possibly different size in order to see how they interact. To this end, we take the initial state to be of the form $\ket{\Psi_0} = \ket{N_{\ell_{N 0}}, M_{\ell_{M 0}}}$, with $2 \leq M \leq N$. The anharmonicity manifold $\mathcal{A}$ contains trivially the states $\ket{N_{\ell_N}, M_{\ell_M}}$, where $\ell_N, \ell_M = 1, \ldots, L$ and $\ell_M \neq \ell_N$. Unlike in the previous examples, however, it is now possible that there are also other states which share the same anharmonicity. This is interesting since it means that collisions between quasiparticles may, at least in principle, actually break them and produce either bare bosons or different kinds of quasiparticles. Having said that, in most cases, conservation of energy prevents this process completely. And even in the rare occasions mixing between the nontrivial and trivial states does happen, it is usually quite weak a phenomenon. For more details, see App.~\ref{app:two_stacks}. In the following discussion, we ignore the nontrivial states altogether.

Like above, the effective Hamiltonian describes interacting two-body physics. In the case of equal-size stacks, the noninteracting part is simply given by $\ham_{\tilde{J}} + \ham_{\tilde{\omega}}$, cf.\ Eq.\ \eqref{eq:effective_hamiltonian}. If $M < N$, we need separate single-quasiparticle Hamiltonians for both of the quasiparticles because the parameters depend on the size. The free motion is again heavily altered by interactions.

\begin{figure}
    \centering
    \includegraphics[width=\columnwidth]{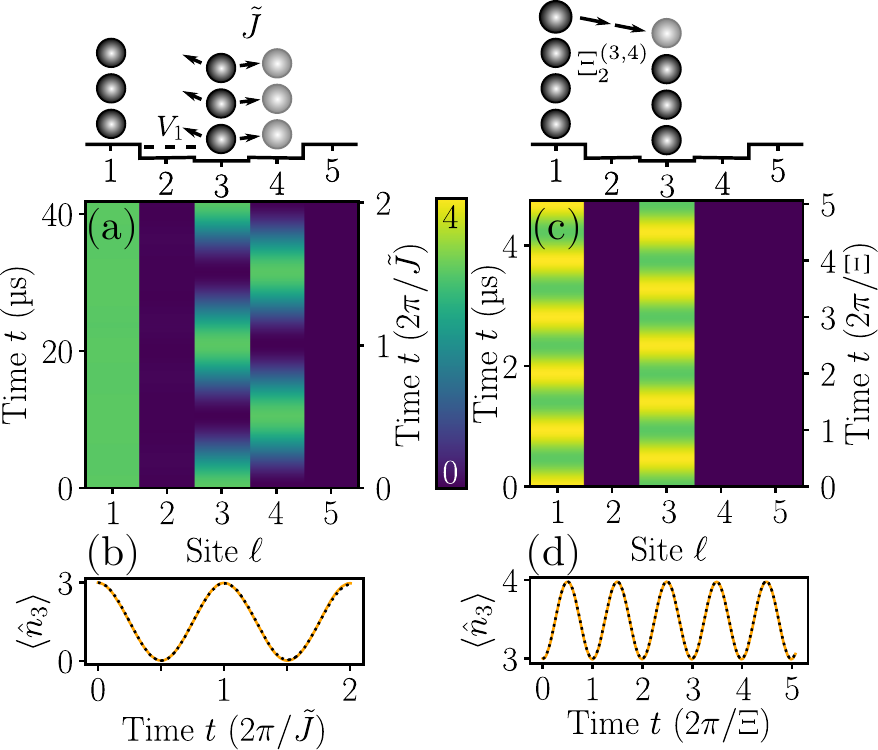}
    \caption{Long-range quasiparticle-quasiparticle repulsion and exchange of distinguishable quasiparticles within a bound pair. (a, c) The local occupations $\langle \hat{n}_\ell \rangle$ as a function of the site index $\ell$ and time $t$. (b, d) The occupation of the site $\ell = 3$ as a function of time $t$, simulated using the full Hamiltonian of Eq.\ \eqref{eq:ham} (solid yellow) and the effective Hamiltonian formed by adding the interaction term of either Eq.\ \eqref{eq:quasiparticle_interaction_same_size} (b) or Eq.\ \eqref{eq:quasiparticle_exchange} (d) to the effective single-quasiparticle Hamiltonian \eqref{eq:effective_hamiltonian} (dotted black). The initial state is $\ket{3_1, 3_3}$ in (a--b) and $\ket{4_1, 3_3}$ in (c--d).}
    \label{fig:two_stacks}
\end{figure}

Assuming first that $M = N$, the quasiparticles interact via the term [cf.\ Eq.\ (\ref{eq:nearest_neighbor_interaction})]
\begin{equation}\label{eq:quasiparticle_interaction_same_size}
    \ham_V / \hbar = \sum_{\ell_1 = 1}^L \sum_{\ell_2 \neq \ell_1} V_{|\ell_2 - \ell_1|} \hat{\nu}_{\ell_1} \hat{\nu}_{\ell_2},
\end{equation}
where the interaction strength is given by
\begin{equation}
    V_\ell = \frac{2 N^3}{(N - 1)^{2 \ell - 1}} \left(1 - \frac{\ell - 1}{N} \frac{2 N - 1}{2 N - 3} \right) \left(\frac{J}{U}\right)^{2 \ell - 1} J.
\end{equation}
Unlike the attractive interaction $\ham_U$ between the bare bosons, $\ham_V$ is \emph{nonlocal}, and so the interaction between quasiparticles can be of longer range. A qualitative picture of the dynamics is simple to paint. The hopping term tries to move a quasiparticle to an adjacent site, but this is only possible if it is strong enough to overcome not only the local potential energy barrier but also the mutual interaction between the quasiparticles. Thus, in addition to being restricted individually by the boundary effects as described in Sec.\ \ref{sec:single_stack}, quasiparticles approaching each other from distance never get closer than $\lceil N/2 \rceil$ sites from one another, whereas for initial separations no more than $\lceil N/2 \rceil - 1$ sites, they form an immobile bound pair. This behavior is again in accordance with numerics as depicted in Fig.\ \ref{fig:two_stacks}(a).

If $M = N - 1$, the interaction term $\ham_V$ of Eq.\ (\ref{eq:quasiparticle_interaction_same_size}) is instead replaced by the long-range exchange Hamiltonian [cf.\ Eq.\ (\ref{eq:exchange_stack+boson})]
\begin{equation}\label{eq:quasiparticle_exchange}
    \ham_\Xi / \hbar = \sum_{\ell_1 = 1}^L \sum_{\ell_2 \neq \ell_1} \Xi_{|\ell_2 - \ell_1|} \hat{\beta}^\dagger_{\ell_2} \hat{\alpha}^\dagger_{\ell_1} \hat{\beta}_{\ell_1} \hat{\alpha}_{\ell_2},
\end{equation}
where $\hat{\alpha}_\ell^\dagger$ and $\hat{\beta}_\ell^\dagger$ are the creation operators for the $N$-boson and $M$-boson quasiparticles, respectively, and
\begin{equation}
    \Xi_\ell = (-1)^{\ell - 1} \frac{N}{(N - 1)^{\ell - 1}} \left(\frac{J}{U}\right)^{\ell - 1} J
\end{equation}
is the exchange rate. Compared to the case of $M = N$ above, the qualitative dynamics here differs in two ways. First, the range of the repulsive interaction between the quasiparticles is now increased, preventing approaching particles from getting closer than $N - 1$ sites from one another. The critical range for bound-state formation is correspondingly increased to $N - 2$ sites. As a new phenomenon, we can observe exchange oscillations within these otherwise stationary bound pairs, that is, the quasiparticles change positions at rate given by $\Xi_\ell$. This is shown in Fig.\ \ref{fig:two_stacks}(b).

As the size difference $N - M$ further increases, both types of interactions discussed above are present in the effective Hamiltonian, but with modified strengths $V_\ell / U \sim (J/U)^{2 \ell}$ and $\Xi_\ell / U \sim (J/U)^{(N - M) \ell}$. Qualitatively speaking, there is nothing new in the dynamics. A long-range repulsion again prevents colliding quasiparticles from getting closer than $\lceil M / 2 \rceil$ sites away from each other. Conversely, quasiparticles bind together if the initial distance between them is at most $\lceil M / 2 \rceil - 1$ sites. For realistic values of $N$ and $M$, exchange oscillations inside these pairs can only show up when the quasiparticles are at adjacent sites. The corresponding exchange rate is given by
\begin{equation}
    \Xi_1 = (-1)^{N - M - 1} \binom{N}{M} \frac{(N - M)^2}{(N - M)!} \left(\frac{J}{U}\right)^{N - M - 1} J.
\end{equation}
The smallest values of $N$ and $M$ allowing exchange over two sites are $7$ and $5$, respectively, and the resulting dynamics would be of fourth order in $J / U$. In principle, it is also possible to observe motion of a bound pair, but the time scales involved are too long for this to be of practical interest, and we shall therefore not explore it further here.

As a final remark, we want to point out that there are also cases where the dynamics is much more complicated to analyze than in the ones discussed above due to the different effects competing with each other. For example, if $M = 2$, $N = 3$, and the initial separation between the quasiparticles is $|\ell_{N 0} - \ell_{M 0}| = 2$, one can observe effects arising from boundaries, exchange, and effective hopping simultaneously, with none clearly dominating the others.

Both of the interaction terms can once again be derived by studying the virtual boson hops. Consider first $\ham_V$. It actually has similar origins as $\ham_{\tilde{\omega}}$. To see this, let us assume that $N - M \neq 1$ and examine the states $\ket{N_{\ell_0}, M_{\ell_0 + \ell}}$ and $\ket{N_{\ell_0}, M_{\ell_0 + \ell + 1}}$ at order $2 \ell$. Forgetting the possible boundary effects which are already taken into consideration, there are a total of $\binom{2 \ell}{\ell}$ trajectories setting these apart. Namely, the ones where lone bosons from each of the stacks first move $m$ and $\ell - m$ sites towards one another, respectively, with $m = 0, \ldots, \ell$, and then back where they originated from. The reason the two states are energetically distinct is that in the case of the former, the bosons actually meet at site $\ell_0 + m$, while in the case of the latter, they are always located at different sites. The leading-order energy difference $V_\ell$ can, in fact, be calculated in closed form for any $N$ and $M$, see App.~\ref{app:two_stacks} for details.

The exchange term is more straightforward to understand. If we take the state $\ket{N_{\ell_0}, M_{\ell_0 + \ell}}$ and move $N - M$ bosons from $\ell_0$ to $\ell_0 + \ell$, we end up with $\ket{N_{\ell_0 + \ell}, M_{\ell_0}}$, establishing a coupling between the two. For general $N - M$, however, the number of trajectories is much larger than above, and calculating the coupling constant $\Xi_\ell$ analytically is therefore tricky unless either $N - M = 1$ or $\ell = 1$. Luckily, these are practically the only relevant instances for us.

\section{\replaced{Disorder and two-dimensional arrays}{Extending to two dimensions}} \label{sec:2D}
\subsection{Disorder in transmon parameters}
\begin{figure}
    \centering
    \includegraphics[width=\columnwidth]{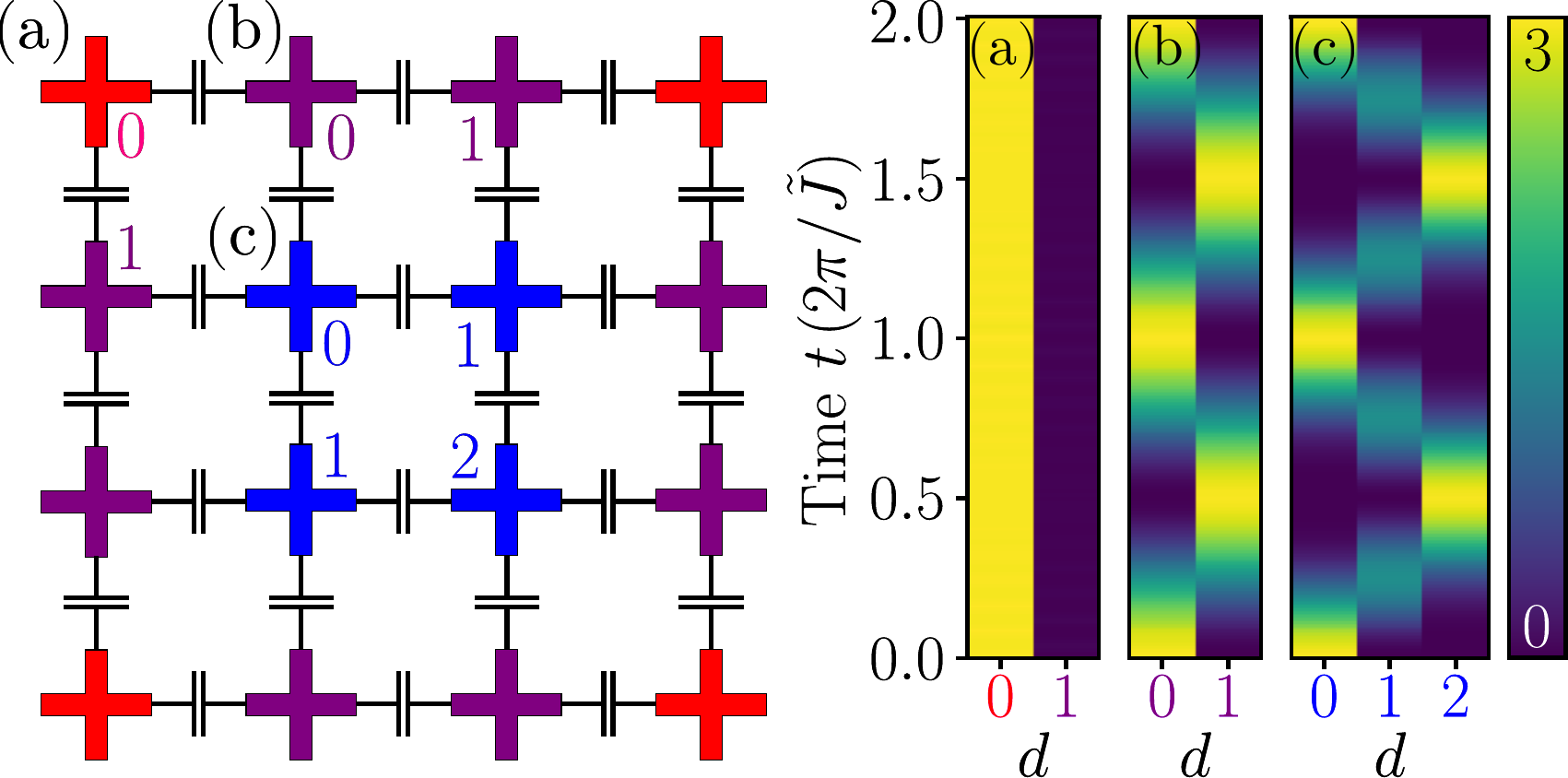}
    \caption{Two-dimensional boundary effects for a single quasiparticle initially located at (a) a corner $\ket{3_{1, 1}}$, (b) an edge $\ket{3_{1, 2}}$, and (c) a non-edge site $\ket{3_{2, 2}}$ of a $4 \times 4$ transmon array. (a--c) The total occupation $\langle \hat n_d \rangle=\sum_{ij}^d\langle \hat n_{ij}\rangle$ of all the sites which are a Manhattan distance $d$ away from the initial location as a function of time~$t$.}
    \label{fig:2D}
\end{figure}
\added{
Until now, we have studied an ideal system, ignoring the inevitable disorder in the model parameters stemming from the imperfections in real transmon devices. Following the discussion of Sec.\ \ref{sec:intro2}, we shall here concentrate on the effects of variations in the single-transmon parameters $\omega_\ell$ and $U_\ell$, and thus take the disorder Hamiltonian to be $\ham_D = \ham_{\delta \omega} + \ham_{\delta U}$. Note that with random $\delta \omega_\ell$ and $\delta U_\ell$, the degeneracy of the states within a given anharmonicity manifold is broken by $\ham_D$. The important question is whether or not this is enough to affect the dynamical phenomenon under examination in a significant manner. Details on the general analysis can be found in App.~\ref{app:theory_disorder}.




In the case of a single stack of bosons, we find (see App.~\ref{app:single_stack}) that it is actually possible to compensate the disorder in $U_\ell$ by adjusting $\omega_\ell$. In fact, if $\omega_\ell = (N - 1) U_\ell / 2$, the disorder Hamiltonian $\ham_D$ vanishes within the anharmonicity manifold spanned by the states $\ket{N_\ell}$. Nevertheless, even with perfect control of $\omega_\ell$, one cannot get rid of the disorder in $U_\ell$ completely. This is because virtual boson hops can take us outside the initial manifold, and there the number of bosons at any given site is no longer guaranteed to be either $N$ or zero. Remember, however, that each virtual jump involved adds a factor of $J/U$ to the weight of the process, therefore effectively reducing the disorder strength. Putting things together, we can define an effective disorder strength $D = \max\{D_\omega, (J/U)^2 D_U \}$, where we understand $D_\omega$ in a slightly wider sense to mean the precision with which the on-site energies can be controlled (without flux tuning, $D_\omega$ is just the regular manufacturing disorder in $\omega_\ell$). As a rough rule of thumb, the motion of the stack is now frozen if $D \gg \tilde{J}$, heavily modified if $D \sim \tilde{J}$, and only slightly changed if $D \ll \tilde{J}$.

The above can be readily applied in the case of two equal-sized stacks. Note that disorder favors bound pairs, and so being able to observe motion is the more challenging problem. If the second stack is of different size, or we add just a single boson, a bit more care is needed since the disorder in $U_\ell$ cannot be entirely eliminated in the whole of the initial anharmonicity manifold simply by adjusting $\omega_\ell$. In general, the effective disorder strength is then increased to $D = \max\{D_\omega, D_U \}$. However, in the phenomena we have concentrated on in this paper -- tunneling and exchange -- one quasiparticle is always either still or moves only between two sites. This restores our ability to control the disorder in $U_\ell$ through $\omega_\ell$, and thus $D = \max\{D_\omega, (J/U)^2 D_U \}$, see Apps.~\ref{app:stack+boson}~and~\ref{app:two_stacks} for details. Again, $2 \pi / D$ gives the critical time scale of observable dynamics.

State-of-the-art transmon arrays~\cite{Arute19, ma_dissipatively_2019, braumuller_probing_2021, Karamlou22, Gong21_walks} exhibit anharmonicity disorder of strength $D_U / 2\pi \approx $~\SIrange{1}{10}{\mega\hertz}. On the other hand, the on-site tuning precision is at best limited to $D_\omega / 2\pi \sim \SI{100}{\kilo\hertz}$. This means that even with quite large a hopping rate of $J/U \sim 1/10$, the effective disorder strength is, at least in the cases considered above, given by $D = D_\omega$. Thus, disorder currently sets the upper bound of $\sim \SI{10}{\micro\second}$ to experimentally observable phenomena.
}

\subsection{Generalization to two-dimensional lattices}
The above results regarding the effective dynamics of transmon chains via high-order perturbation theory are readily generalizable to two-dimensional lattices, although with some differences to be kept in mind. The effective hopping Hamiltonian of the quasiparticle simply inherits the geometry of the lattice, that is, it is obtained from the hopping Hamiltonian $\ham_J$ with the replacements $J \to \tilde{J}$, $\hat{a}_\ell \to \hat{\alpha}_\ell$. Since the trajectory determining the effective hopping frequency includes just moving bosons from one site to an adjacent one, the value of $\tilde{J}$ does not depend on the dimensionality of the lattice and is thus always given by Eq.~\eqref{eq:tilde_J}. Distinct couplings between the sites can also be incorporated into the model straightforwardly by setting $J \to J_{\ell_1 \ell_2}$.

In a two-dimensional array, the boundary effects are similar to a 1D chain, but more possibilities arise from the increased dimensionality. For example, in a rectangular array, there are now two distinct groups of boundary sites, namely, the corners and the edges. This stems from the different number of neighbors each site possesses. Thus, putting a stack of $N \ge 3$ particles initially into a corner site shows no dynamics at all [Fig.~\ref{fig:2D}(a)], whereas taking the initial site to be one of the edge sites allows movement of the stack within that edge [Fig.~\ref{fig:2D}(b)]. Similarly as with the one dimensional case, the dynamics within the sites that are neither edges nor corners is practically free, and described by the effective hopping rates $\tilde{J}$ of Eq.~\eqref{eq:tilde_J} [Fig.~\ref{fig:2D}(c)].

Finally, when considering the interaction between two quasiparticles, the strength is determined by the geodesic (Manhattan) distance on the underlying graph. The dynamics of an unbound pair is also more complex due to the fact that the quasiparticles can now move past each other.

\section{Conclusions} \label{sec:conc}
In this work, we studied unitary dynamics of weakly-coupled transmon arrays, concentrating specifically on the phenomena brought about by the higher excited states of the individual transmons. The key observation considering many-body dynamics beyond the qubit approximation or, equivalently, beyond the hard-core boson model, was the approximate conservation of the interaction energy stemming from the anharmonicity of the transmons. Based on this, we were able to resolve the dynamics using high-order degenerate perturbation theory, whose accuracy we then benchmarked with exact numerics. 

The main results demonstrated various many-body effects, which we presented in closed form using effective Hamiltonians. For example, bosons initially stacked onto the same site behave as a single quasiparticle whose effective hopping frequency depends exponentially on the boson number $N$. In other words, a highly-excited state of a transmon does not disintegrate into several less-excited states, but instead just moves from one transmon to another. The quasiparticles also experience effective off-site interactions with other quasiparticles, individual bosons\deleted{(single excitations)}, and the edges and corners of the arrays. The presented approximation significantly reduces the dimensionality of the Hilbert space since the dynamics and the energy levels can be solved independently within each anharmonicity manifold. Most importantly, the dynamics generated by the closed-form effective Hamiltonians \replaced{were}{was} found to be accurate well up to the time scales relevant to any given subspace. This allows us to explore the largely ignored portion of the Hilbert space of a transmon array going beyond the hard-core boson model. \added{For practical observation of the presented phenomena, ability to prepare and measure highly excited transmon states, that is, local boson occupation density, at high fidelity is an essential requirement. In general, our results are readily applicable also to other similar systems, such as cold atoms in optical lattices~\cite{Bloch08, Schafer20}, which are modeled by the Bose-Hubbard model with low parameter disorder and operated in the limit $J/U\ll 1$ where interactions dominate over the hopping rate.}

Our focus here was on unitary dynamics. Naturally, dissipation and dephasing of transmons will generate notable effects whose detailed numerical and analytical consideration is a subject of \replaced{future}{current} research. \deleted{ignoring any non-unitary effects caused by, for example, dissipation and dephasing processes or measurements.} \added{To elucidate the experimental feasibility of quasiparticle dynamics in transmon arrays, let us focus here on a simple time scale analysis. Treating a multi-level transmon ideally as a harmonic oscillator, a quasiparticle made of $N$ photons has a decay rate of $\Gamma^{(N)} = N \Gamma^{(1)}$ due to bosonic enhancement. Here, $\Gamma^{(1)}=T_1^{-1}$ is the single-particle decay rate. Similarly, the dephasing rate between the consecutive states $\ket{N+1}$ and $\ket{N}$ would scale as $\Gamma_2^{(N, N-1)} = \Gamma^{(N-1)} / 2 + \Gamma_2^{(0, 1)}$, where $\Gamma_2^{(0, 1)} = T_2^{-1}$ refers to the dephasing rate in the qubit subspace. With realistic transmons~\cite{Koch07, Peterer15, Blok21}, the scaling of the decay rate follows quite well that of an ideal harmonic oscillator, but the coherence of the excited states is reduced due to their enhanced susceptibility to charge noise. In state-of-the-art quantum simulation devices~\cite{Arute19, ma_dissipatively_2019, braumuller_probing_2021, Karamlou22, Gong21_walks, Zhang22}, the qubit-subspace dissipation and dephasing rates are given by $T_1 = $~\SIrange{10}{30}{\micro\second} and $T_2 = $~\SIrange{1}{3}{\micro\second}, respectively, resulting in the estimates of $1 / \Gamma^{(N)} \gtrsim $~\SIrange{3}{10}{\micro\second} and $1 / \Gamma_2^{(N, N-1)} \lesssim $~\SIrange{1}{3}{\micro\second} for the higher-excited-state dissipation and dephasing times when $N \leq 3$.  The effective rates for the quasiparticle dynamics scale as $(J/U)^N$ implying that with realistic transmon parameters $|J| / 2\pi \approx $~\SIrange{10}{40}{\mega\hertz} and $U /2\pi = $~\SIrange{200}{300}{\mega\hertz}, the time scales of the many-body dynamics achieved (see~Figs.~\ref{fig:single_stack}-~\ref{fig:two_stacks}) are of the order of~\SIrange{0.3}{10}{\micro\second}, rendering experimental realization of the presented phenomena possible using state-of-the-art transmon arrays. Furthermore, quantum-information-oriented transmon systems~\cite{Krinner22, Blok21} have recently reported decay and dephasing times as high as $T_{1,2} \approx $~\SI{70}{\micro\second}, which would already yield an ample window to observe the quasiparticle dynamics.}

Various potential applications also arise. For example, transmon populations beyond the qubit subspace can be measured with high fidelity either using a direct dispersive circuitQED readout~\cite{Blok21} or by supplementing it with conditional pulses~\cite{Peterer15}. This combination of having periodic high-fidelity measurements and rich many-body dynamics makes transmon arrays a promising experimental platform for realizing measurement-induced entanglement phase transitions~\cite{Li18, Skinner19, Chan19, Potter21, Minnich22}, complementing ion traps~\cite{Noel21, Czischek21}. On the other hand, for the sake of simplicity, the array geometries in this work have been kept quite basic. An interesting direction to extend the presented analysis is towards more complex geometries, such as Kagome~\cite{Koch10} or non-Euclidian~\cite{Kollar19}, for instance to probe intriguing flat band physics. Furthermore, versatile many-body dynamics provides a good basis for studying non-equilibrium many-body dynamics, such as dynamical quantum phase transitions~\cite{Heyl18}. We also expect that our results and concepts on quasiparticle dynamics can be useful in understanding and solving significant design challenges in quantum processor architectures similarly as the concepts of many-body localization has been applied to the protection-operation dilemma of quantum computing with transmon arrays~\cite{Berke21}. 

\section*{Acknowledgements}
We are grateful to Oksana Busel, Steven Girvin, and Tuure Orell for useful discussions. We acknowledge financial support from the Emil Aaltonen foundation, the Kvantum Institute of the University of Oulu, and the Academy of Finland under Grants Nos.~316619, 320086, and 346035.

\appendix
\section{High-order degenerate perturbation theory}\label{app:theory}
\subsection{Projected Schrödinger equation}
Let us consider a general Hamiltonian $\ham = \ham_0 + \ham_1$ consisting of two parts, a \enquote{trivial} part $\ham_0$ whose eigenproblem we can solve, and some perturbation $\ham_1$. Let $\mathcal{E}_0$ be the eigenspace of $\ham_0$ related to some eigenvalue $E_0$, that is, $\mathcal{E}_0$ is the space spanned by all the states $\ket{E_0}$ satisfying $\ham_0 \ket{E_0} = E_0 \ket{E_0}$. Finally, let $\hat{P}_0$ be the projection operator to $\mathcal{E}_0$ and $\hat{Q}_0 = \hat{I} - \hat{P}_0$ the projection operator to the complement $\mathcal{E}_0^c$ of $\mathcal{E}_0$.

Projecting the time-independent Schrödinger equation $\ham \ket{E} = E \ket{E}$ of the full Hamiltonian into $\mathcal{E}_0$ and $\mathcal{E}_0^c$, respectively, and using the identities $\hat{I} = \hat{P}_0 + \hat{Q}_0$, $\ham_0 \hat{P}_0 = \hat{P}_0 \ham_0 = E_0 \hat{P}_0$, we obtain
\begin{align}
    \hat{P}_0 \ham_1 \hat{P}_0 \ket{E} + \hat{P}_0 \ham_1 \hat{Q}_0 \ket{E} &= (E - E_0) \hat{P}_0 \ket{E}, \\
    \hat{Q}_0 (E - \ham_0 - \ham_1) \hat{Q}_0 \ket{E} &= \hat{Q}_0 \ham_1 \hat{P}_0 \ket{E}.
\end{align}
Solving the second equation for $\hat{Q}_0 \ket{E}$ (we assume that the effect of $\ham_1$ is sufficiently small to avoid any divergence issues) and using this in the first equation gives us
\begin{align}
    &\hat{P}_0 \ham_1 \left\{\hat{I} + \left[\hat{Q}_0 (E - \ham_0 - \ham_1) \hat{Q}_0 \right]^{-1} \hat{Q}_0 \ham_1 \right\} \hat{P}_0 \ket{E} \nonumber \\
    &\hspace{27pt}= (E - E_0) \hat{P}_0 \ket{E}, \label{eq:app_schr_0} \\
    &\hat{Q}_0 \ket{E} = \left[\hat{Q}_0 (E - \ham_0 - \ham_1) \hat{Q}_0 \right]^{-1} \hat{Q}_0 \ham_1 \hat{P}_0 \ket{E}. \label{eq:app_schr_1}
\end{align}
The first equation here is a generalized eigenvalue equation (note that $E$ appears also on the left-hand side) determining the projection of the eigenstate within $\mathcal{E}_0$. After solving for $\hat{P}_0 \ket{E}$ and $E$, the second equation can then be used to straightforwardly calculate the projection of the eigenstate within the rest of the Hilbert space.

As a final step, we apply the general formula
\begin{equation}
    (\hat{M} + \hat{N})^{-1} = (\hat{I} + \hat{M}^{-1} \hat{N})^{-1} \hat{M}^{-1} = \sum_{m = 0}^{\infty} (\hat{M}^{-1} \hat{N})^m \hat{M}^{-1}
\end{equation}
to the inverse operators appearing in Eqs.\ (\ref{eq:app_schr_0}) and (\ref{eq:app_schr_1}), yielding
\begin{align}
    \sum_{m = 0}^{\infty} \hat{A}_m(E) \hat{P}_0 \ket{E} &= (E - E_0) \hat{P}_0 \ket{E}, \label{eq:app_schr_02} \\
    \hat{Q}_0 \ket{E} &= \sum_{m = 1}^{\infty} \hat{B}_m(E) \hat{P}_0 \ket{E}, \label{eq:app_schr_12}
\end{align}
where the $\hat{A}_m(E)$ and $\hat{B}_m(E)$ operators are
\begin{align}
    \hat{B}_m(E) &= \left\{\left[\hat{Q}_0 (E - \ham_0) \hat{Q}_0 \right]^{-1} \ham_1 \right\}^{m} \hat{P}_0, \\
    \hat{A}_m(E) &= \hat{P}_0 \ham_1 \hat{B}_m.
\end{align}
Equation (\ref{eq:proj_schrödinger}) then follows trivially from Eq.~(\ref{eq:app_schr_02}).

\subsection{Degenerate perturbation theory}
Let us then discuss in general terms how Eq.\ (\ref{eq:proj_schrödinger}) can be solved within a given anharmonicity manifold $\mathcal{A}$. Expanding the states $\ket{E_\mathcal{A}}$ and the energies $E$ in powers of $J/U$ as $\ket{E_\mathcal{A}} = \ket{E_\mathcal{A}^{(0)}} + \ket{E_\mathcal{A}^{(1)}} + \ldots$ and $E = E^{(0)} + E^{(1)} + \ldots$, respectively, and collecting terms of equal order together yields
\begin{equation}\label{eq:degenerate_perturbation_theory_full}
\sum_{j = 0}^{n} \ham_\mathcal{A}^{(n - j)} \ket{E_\mathcal{A}^{(j)}} = \sum_{j = 0}^{n} E^{(n - j)} \ket{E_\mathcal{A}^{(j)}}
\end{equation}
for $n = 0, 1, 2, \ldots$. Here $\ham_\mathcal{A}^{(k)}$ is used to denote the $k$th-order term of $\ham_\mathcal{A}(E)$. Note that this is not the same as $\hat{K}_k$ which, while explicitly of $k$th order, still depends on the energy. As usual, we need to proceed order by order. In this paper, we are not interested in the higher-order states, and so we only concentrate here on solving for $\ket{E_\mathcal{A}^{(0)}}$.

At zeroth order, we obtain trivially $E^{(0)} = \hbar U A$, that is, all the states share the common zeroth-order energy. Nothing is revealed about the states $\ket{E_\mathcal{A}^{(0)}}$, all we know at this point is that they lie within $\mathcal{A}$.

At first order, we have
\begin{equation}\label{eq:degenerate_perturbation_theory_first_order}
    \ham_\mathcal{A}^{(1)} \ket{E_\mathcal{A}^{(0)}} =  E^{(1)} \ket{E_\mathcal{A}^{(0)}}.
\end{equation}
We therefore see that the first-order energies are given by the eigenvalues of the $d \times d$ matrix $\ham_\mathcal{A}^{(1)}$, where $d$ is the dimension of $\mathcal{A}$. If the spectrum of $\ham_\mathcal{A}^{(1)}$ happens to be non-degenerate, as is the case in most textbook examples, then the eigenstates of $\ham_\mathcal{A}^{(1)}$ are the proper zeroth-order eigenstates $\ket{E_\mathcal{A}^{(0)}}$ and no further analysis is needed. This is because we can now uniquely tell apart the states based on their energies, and thus taking the limit $J/U \to 0$ is no longer problematic.

In the situations we study, however, degeneracy is always present also at first order. In this case, just like the unperturbed Hamiltonian $\ham_U$ splits the whole Hilbert space into different anharmonicity manifolds according to its eigenvalues, the effective first-order Hamiltonian $\ham_\mathcal{A}^{(1)}$ splits each anharmonicity manifold further into (possibly) smaller subspaces according to its eigenvalues. Higher-order analysis can then be performed separately in each of these instead of considering the whole $\mathcal{A}$ at once since we know that each state $\ket{E_\mathcal{A}^{(0)}}$ always belongs to exactly one such subspace (the non-degenerate case discussed above is simply a special case of this, with $d$ one-dimensional subspaces).

At second order, we have
\begin{equation}\label{eq:degenerate_perturbation_theory_second_order_full}
    \ham_\mathcal{A}^{(2)} \ket{E_\mathcal{A}^{(0)}} + \ham_\mathcal{A}^{(1)} \ket{E_\mathcal{A}^{(1)}} = E^{(2)} \ket{E_\mathcal{A}^{(0)}} + E^{(1)} \ket{E_\mathcal{A}^{(1)}}.
\end{equation}
Note the appearance of the first-order states on both sides of the equation. Although they might seem problematic, we can get rid of them quite easily. After picking one of the first-order energies $E^{(1)}$ to concentrate on, we can project Eq.\ (\ref{eq:degenerate_perturbation_theory_second_order_full}) to the corresponding eigenspace, as we discussed above. But since Eq.\ (\ref{eq:degenerate_perturbation_theory_first_order}) holds for all states within the eigenspace, we have the identity $\ham_\mathcal{A}^{(1)} = E^{(1)} \hat{I}$, and thus the terms involving the first-order state cancel out. We are therefore left with
\begin{equation}\label{eq:degenerate_perturbation_theory_second_order}
    \ham_\mathcal{A}^{(2)} \ket{E_\mathcal{A}^{(0)}} = E^{(2)} \ket{E_\mathcal{A}^{(0)}}.
\end{equation}
This is again an eigenvalue equation, and exactly the same arguments that were presented at first order hold also here.

By following this procedure of always projecting to one of the still degenerate subspaces, Eq.\ (\ref{eq:degenerate_perturbation_theory_full}) simplifies to
\begin{equation}\label{eq:degenerate_perturbation_theory}
    \ham_\mathcal{A}^{(n)} \ket{E_\mathcal{A}^{(0)}} = E^{(n)} \ket{E_\mathcal{A}^{(0)}},
\end{equation}
which is, at every order, an eigenvalue equation. We continue until either all the degeneracies are lifted, in which case we know exactly all the states $\ket{E_\mathcal{A}^{(0)}}$, or we reach some predetermined limit $n_{\max}$ above which it is not necessary to go due to the slowness of the resulting dynamics.

Expressed in terms of the operators $\hat{K}_m$ defined in Eq.~(\ref{eq:K_n}), we can write
\begin{equation}\label{eq:Hamiltonian_nth_order}
    \ham_\mathcal{A}^{(n)} = \sum_{m = 1}^{n} \hat{K}_m^{(n - m)}
\end{equation}
for $n \geq 1$. Here the notation $\hat{K}_m^{(k)}$ again means that we expand $\hat{K}_m(E)$ in energy and take the $k$th term, so that $\hat{K}_m^{(k)}$ is of $(m + k)$th order. For example, since $\hat{K}_1(E) = \hat{P}_\mathcal{A} \ham_J \hat{P}_\mathcal{A}$ is independent of $E$, only $\hat{K}_1^{(0)}$ is nonzero. Importantly, no matter the value of $k$, if $\hat{K}_m^{(k)}$ is nonzero, it contains exactly $m$ hopping Hamiltonians $\ham_J$. This implies that $\ham_\mathcal{A}^{(n)}$ is capable of performing at most $n$ single-boson hops.

\subsection{Effect of disorder}\label{app:theory_disorder}
\added{
How does disorder in transmon parameters affect the zeroth-order eigenstates and the dynamics? To answer this, we need to add the disorder Hamiltonians $\ham_{\delta \omega} = \hbar \sum_\ell \delta \omega_\ell \hat{n}_\ell$ and $\ham_{\delta U} = - \hbar \sum_\ell \delta U_\ell \hat{n}_\ell (\hat{n}_\ell - 1) / 2$ alongside the hopping Hamiltonian $\ham_J$ in the definition of $\hat{K}_m$ in Eq.\ (\ref{eq:K_n}). Here $\delta \omega_\ell = \omega_\ell - \omega$ and $\delta U_\ell = U_\ell - U$ are the deviations of the on-site energies and the interaction energies from the constant values $\omega$ and $U$ at different sites $\ell$, respectively. We ignore here the disorder in $J$ since it can be thought of as a small correction to the already small perturbation parameter $J$. The total disorder Hamiltonian is therefore given by $\ham_D = \ham_{\delta \omega} + \ham_{\delta U}$.}

\added{Replacing $\ham_J \to \ham_J + \ham_D$ in Eq.\ (\ref{eq:K_n}) and using the fact that $\ham_D$ commutes with $\ham_U$, we obtain}
\begin{align}
\hat{K}_1(E) &= \hat{P}_\mathcal{A} \ham_J \hat{P}_\mathcal{A} + \hat{P}_\mathcal{A} \ham_D \hat{P}_\mathcal{A}, \label{eq:K_1_disorder} \\
\hat{K}_m(E) &= \hat{P}_\mathcal{A} \ham_J [\hat{W}(E) (\ham_J + \ham_D) ]^{m - 2} \hat{W}(E) \ham_J \hat{P}_\mathcal{A}, \label{eq:K_m_disorder}
\end{align}
\added{where $m = 2, 3, \ldots$ and the definition of $\hat{W}(E)$ remains unchanged from Eq.\ (\ref{eq:W}). Here it is important to note the two different manifestations of disorder. First, $\hat{K}_1$ depends solely on the disorder within the anharmonicity manifold $\mathcal{A}$. All the other operators $\hat{K}_m$, on the other hand, contain only the projection of $\ham_D$ outside of $\mathcal{A}$ (due to the projectors contained in the weight operator $\hat{W}$). It is therefore natural to define two separate disorder strengths, $D_\mathcal{A}$ and $D'_\mathcal{A}$, so that $\hat{P}_\mathcal{A} \ham_D \hat{P}_\mathcal{A} \sim \hbar D_\mathcal{A}$ and $\hat{Q}_\mathcal{A} \ham_D \hat{Q}_\mathcal{A} \sim \hbar D'_\mathcal{A}$.}

\added{Let us then assume that $D_\mathcal{A} / U \sim (J / U)^n$ and $D'_\mathcal{A} / U \sim (J / U)^{n'}$ for some $n, n' \in \mathbb{N}$. We allow for $n \geq n'$ since it might be possible to reduce the disorder within $\mathcal{A}$ by tuning the on-site energies. Performing the perturbation analysis as above, we see that $\hat{P}_\mathcal{A} \ham_D \hat{P}_\mathcal{A}$ makes its first appearance at $n$th order (through $\hat{K}_1$), while $\hat{Q}_\mathcal{A} \ham_D \hat{Q}_\mathcal{A}$ first appears at order $n' + 2$ (through $\hat{K}_3$). Due to the irregular structure of the on-site energies and interactions, all the degeneracy is thus broken (at least almost surely) at order $\min\{n, n' + 2 \}$. }

\added{The above discussion invites us to define the effective disorder strength $D^\mathrm{eff}_\mathcal{A} = \max \{ D_\mathcal{A}, D'_\mathcal{A} (J/U)^2 \}$. Roughly speaking, all the dynamical effects of the pure system occurring at time scales slower than $\sim 2 \pi / D^\mathrm{eff}_\mathcal{A}$ are wiped away by the disorder. Dynamics at the time scale $\sim 2 \pi / D^\mathrm{eff}_\mathcal{A}$ is expected to be significantly modified, while faster dynamics should remain more or less the same.}


\subsection{One-boson problem}
We will refer to the dynamics of a lone boson in several occasions, and it is therefore good to briefly recall some of the properties of the one-boson problem.

Let us consider a chain of length $L$ with constant nearest-neighbor hopping frequency $J$ and on-site energies $\hbar \omega_\ell$. The interaction term $\ham_U$ is now identically zero since there are no other bosons to interact with. The Hamiltonian is therefore given by
\begin{equation}
    \ham / \hbar = \sum_{\ell = 1}^{L} \omega_\ell \hat{n}_\ell + J \sum_{\ell = 1}^{L - 1} (\hat{a}_{\ell + 1}^\dagger \hat{a}_\ell + \hat{a}_\ell^\dagger \hat{a}_{\ell + 1}).
\end{equation}
Written in matrix form in the basis $\ket{1_\ell}$, $\ell = 1, \ldots, L$, we have
\begin{equation}
    H / \hbar =
    \begin{pmatrix}
        \omega_1 & J & & & & \\
        J & \omega_2 & J & & & \\
        & J & \omega_3 & J & & \\
        & & \ddots & \ddots & \ddots & \\
        & & & J & \omega_{L - 1} & J \\
        & & & & J & \omega_L
    \end{pmatrix}.
\end{equation}
This is a symmetric tridiagonal matrix with nonzero off-diagonal elements, and thus the spectrum is always nondegenerate \cite{parlett1980}. For general values of $\omega_\ell$, the eigenvalues and eigenstates have no closed-form analytical expressions.

In the special case of constant $\omega_\ell = \omega$, the Hamiltonian is a tridiagonal Toeplitz matrix whose eigenproblem is analytically solvable \cite{losonczi1992}. The eigenvalues are given by
\begin{equation}\label{eq:single-particle_energies}
    \varepsilon_k / \hbar = \omega + 2 J \cos\left(\frac{\pi k}{L + 1} \right)
\end{equation}
for $k = 1, \ldots, L$, and the corresponding eigenstates by
\begin{equation}\label{eq:single-particle_states}
    \ket{\varepsilon_k} = \sqrt{\frac{2}{L + 1}} \sum_{\ell = 1}^{L} \sin\left(\frac{\pi \ell k}{L + 1} \right) \ket{1_\ell}.
\end{equation}
In the language of operators, defining
\begin{equation}
    \hat{a}_\ell = \sqrt{\frac{2}{L + 1}} \sum_{k = 1}^{L} \sin\left(\frac{\pi \ell k}{L + 1} \right) \hat{c}_k
\end{equation}
turns the Hamiltonian into
\begin{equation}
    \ham / \hbar = \sum_{k = 1}^L \left[\omega + 2 J \cos\left(\frac{\pi k}{L + 1} \right) \right] \hat{c}_k^\dagger \hat{c}^{}_k,
\end{equation}
while the functional form of the total number operator $\hat{N} = \sum_{\ell = 1}^{L} \hat{a}_\ell^\dagger \hat{a}^{}_\ell = \sum_{k = 1}^{L} \hat{c}_k^\dagger \hat{c}^{}_k$ remains unchanged.

\section{Collective motion of a boson stack}\label{app:single_stack}
\subsection{Degenerate states}
The anharmonicity of the initial state $\ket{N_{\ell_0}}$ is $-N(N - 1)/2$. Clearly all the states $\ket{N_\ell}$, $\ell = 1, \ldots, L$, share the same value, but are there any other Fock states $\ket{n_1 n_2 \ldots n_L}$ in this manifold when the total number of bosons is fixed? Well,
\begin{equation}\label{eq:occupation_squares_inequality}
\begin{split}
    N^2 &= \sum_{\ell = 1}^{L} n_\ell \sum_{m = 1}^{L} n_m = \sum_{\ell = 1}^{L} n_\ell^2 + \sum_{\ell = 1}^{L} \sum_{m \neq \ell} n_\ell n_m \\
    &\geq \sum_{\ell = 1}^{L} n_\ell^2,
\end{split}
\end{equation}
and the equality clearly holds only if $n_\ell n_m = 0$ for all $\ell, m = 1, \ldots, L$, $m \neq \ell$, since the $n_\ell$ are nonnegative. But this condition is equivalent to $n_{\ell} = N \delta_{\ell \ell'}$ for some $\ell'$, and so there are no nontrivial states in the anharmonicity manifold.

\subsection{Perturbation analysis}
Let us then apply Eq.\ (\ref{eq:schrödinger_n}) inside the anharmonicity manifold $\mathcal{A} = \mathrm{span} \{\ket{N_\ell} | \ell = 1, \ldots, L \}$, and make use of the representation (\ref{eq:Hamiltonian_nth_order}) of $\ham_\mathcal{A}^{(n)}$. Before any actual calculations, we notice that the states are not coupled until $N$th order: In order for the matrix element $\braket{N_{\ell'} | \hat{K}_m | N_\ell}$ to be nonzero, we need to transfer $N$ bosons from site $\ell$ to site $\ell'$. Since $\hat{K}_m$ contains $m$ hopping Hamiltonians, we need $m$ to be at least $N$ for this to be possible. This means that up to $N$th order, the matrix $\ham_\mathcal{A}^{(n)}$ is diagonal. To simplify the analysis further, we note that the diagonal elements $\braket{N_\ell | \hat{K}_m | N_\ell}$ vanish for odd values of $m$ since moving the bosons around and coming back to the same configuration always requires even number of hops.

The zeroth-order energy is given by $E^{(0)} / \hbar = - U N (N - 1) / 2$. At first order, we find $E^{(1)} = 0$. In fact, at odd orders below $N$ all the energies vanish. This can be seen using induction. In the matrix $\sum_{m = 1}^{n} \hat{K}_{m}^{(n - m)}$ at some odd $n$, all the odd-$m$ terms vanish identically as discussed above. All the even-$m$ terms, on the other hand, need to be expanded to odd order $< n$ in energy. But since all the previous odd-order energies vanish, these terms are all zero, proving our claim.

At second order, we have
\begin{equation}
    \hat{K}_2^{(0)} \ket{E_\mathcal{A}^{(0)}} = E^{(2)} \ket{E_\mathcal{A}^{(0)}}.
\end{equation}
Since all the states had the same first-order energy, this equation is to be solved within the full space $\mathcal{A}$. When the matrix $\hat{K}_2^{(0)} = \hat{P}_\mathcal{A} \ham_J \hat{W}^{(0)} \ham_J \hat{P}_\mathcal{A}$ operates on a basis state $\ket{N_\ell}$, it first generates the Fock states $\ket{(N - 1)_\ell, 1_{\ell \pm 1}}$ which are one hop away, weights them by $\hat{W}^{(0)} = \hat{W}(E^{(0)})$, generates all the possible Fock states which are one hop away from these intermediate states, and finally picks only the ones which are within $\mathcal{A}$. But this shows at once that the edge sites $\ell = 1, L$ are different from all the other ones since one of the sites $\ell \pm 1$ does not belong to the chain. That is, a boson at a boundary site can only move to one direction. This increases the energy of an edge site, leading to
\begin{equation}
    \hat{K}_2^{(0)} \ket{N_\ell} = - \hbar U \left(\frac{J}{U} \right)^2 \frac{2 N}{N - 1} \ket{N_\ell}
\end{equation}
for $\ell = 2, \ldots, L - 1$, and
\begin{equation}
    \hat{K}_2^{(0)} \ket{N_\ell} = - \hbar U \left(\frac{J}{U} \right)^2 \frac{N}{N - 1} \ket{N_\ell}
\end{equation}
for $\ell = 1, L$.

The anharmonicity manifold $\mathcal{A}$ is therefore split into two parts at second order, $\mathrm{span}\{ \ket{N_1}, \ket{N_L} \}$ and $\mathrm{span}\{ \ket{N_2}, \ldots, \ket{N_{L - 1}} \}$. The states $\ket{N_1}$ and $\ket{N_L}$ are $L - 1$ sites away from each other, and therefore the degeneracy between them is not lifted until $[(L - 1) N]$th order, at which they couple together. Since the energy difference is so minute, the time interval required for observing any dynamics between the two states is too long to be of any practical interest. We can therefore treat the states $\ket{N_1}$ and $\ket{N_L}$ as proper zeroth-order eigenstates $\ket{E_\mathcal{A}^{(0)}}$. Thus, if the initial state is localized at the boundary ($\ell_0 = 1, L$), it will stay there and no dynamics is observed at sensible time scales.

Analyzing the space $\mathrm{span}\{ \ket{N_2}, \ldots, \ket{N_{L - 1}} \}$ further, we find that the behavior observed at second order repeats at every even order less than $N$. That is, the two sites closest to the boundaries always have higher energy than the middle ones, and the subspace splits again into two smaller parts, one spanned by the two \enquote{edge} states and the other by the rest. For example, assuming that $N > 6$, fourth order separates the states $\ket{N_2}$ and $\ket{N_{L - 1}}$ from the states $\ket{N_3}, \ldots, \ket{N_{L - 2}}$, while sixth order splits the latter ones into two groups, the states $\ket{N_3}$ and $\ket{N_{L - 2}}$, and the states $\ket{N_4}, \ldots, \ket{N_{L - 3}}$. For sufficiently short chains, the degeneracy can be completely broken before the coupling plays any role (or almost, if there are even number of sites).

The energy difference $\Delta E^{(2 n)}$ between the two sets of states at each order $2 n < N$ is straightforward to calculate since it always stems from a single source: the ability to move one boson from the stack $n$ steps towards the nearest boundary and then back. Thus, if $\ell$ is any site closer to the middle of the chain than the site $n$, we have
\begin{align}
    \Delta E^{(2 n)} &\equiv  \braket{N_n | \ham_\mathcal{A}^{(2 n)} | N_n} - \braket{N_\ell | \ham_\mathcal{A}^{(2 n)} | N_\ell} \notag \\
     &= -\bra{N_\ell} \hbar J \hat{a}_{\ell}^\dagger \hat{a}_{\ell - 1} \prod_{m = 1}^{n - 1} \hat{W}^{(0)} \hbar J \hat{a}_{\ell - m}^\dagger \hat{a}_{\ell - m - 1} \notag \\
     &\hspace{60pt} \times \prod_{m = n}^{1} \hat{W}^{(0)} \hbar J \hat{a}_{\ell - m}^\dagger \hat{a}_{\ell - m + 1} \ket{N_\ell} \notag \\
     &= - (\hbar J)^{2 n} \sqrt{N}^2 [- \hbar U (N - 1)]^{-(2 n - 1)} \notag \\
    &= \frac{\hbar U N}{(N - 1)^{2 n - 1}} \left(\frac{J}{U} \right)^{2 n}.
\end{align}

At $N$th order, there is a coupling between the neighboring states $\ket{N_\ell}$, $\ket{N_{\ell \pm 1}}$. The coupling strength is again straightforward to calculate since the process only involves shifting the bosons individually to the adjacent site. That is,
\begin{equation}
\begin{split}
   \hbar \tilde{J} &\equiv \braket{N_{\ell + 1} | \ham_\mathcal{A}^{(N)} | N_\ell} \\
   &= \braket{N_{\ell + 1} | \hbar J \hat{a}_{\ell + 1}^\dagger \hat{a}_\ell \left[\hat{W}^{(0)} \hbar J \hat{a}_{\ell + 1}^\dagger \hat{a}_\ell \right]^{N - 1} | N_\ell} \\
   &= (\hbar J)^N \prod_{m = 0}^{N - 1} \sqrt{N - m} \sqrt{m + 1} \\
   &\hspace{30pt} \times \prod_{m = 0}^{N - 2} [-\hbar U (m + 1) (N - m - 1)]^{-1} \\
   &= (-1)^{N - 1} \frac{\hbar U N}{(N - 1)!} \left(\frac{J}{U} \right)^{N}.
\end{split}
\end{equation}

If we now write the matrix $\ham_\mathcal{A}^{(N)}$ in the basis of the remaining degenerate states $\ket{N_\ell}$, $\ell = \lceil N/2 \rceil, \ldots, L - \lceil N/2 \rceil + 1$, we obtain
\begin{equation}
    H_\mathcal{A}^{(N)} / \hbar = 
    \begin{pmatrix}
        \omega + \Delta & \tilde{J} & & & & \\
        \tilde{J} & \omega & \tilde{J} & & & \\
        & \tilde{J} & \omega & \tilde{J} & & \\
        & & \ddots & \ddots & \ddots & \\
        & & & \tilde{J} & \omega & \tilde{J} \\
        & & & & \tilde{J} & \omega + \Delta
    \end{pmatrix}.
\end{equation}
Here, $\omega$ is some constant value irrelevant for determining the state while $\Delta = \Delta E^{(N/2)} / \hbar$ if $N$ is even and $\Delta = 0$ if $N$ is odd. But this is simply the Hamiltonian of a single particle in a chain of length $L - 2 \lceil N/2 \rceil + 2$ with hopping frequency $\tilde{J}$ and on-site energies $\omega + \Delta$ at the ends and $\omega$ in the middle, written in the basis $\ket{1_\ell}$. The remaining degeneracy is thus completely lifted.

We have therefore seen that a stack of bosons behaves exactly like a single individual boson in a chain with modified hopping frequency and on-site energies. Using the notation of the main text, we can combine all of the above analysis into the effective Hamiltonian
\begin{equation}
\ham_{\mathrm{eff}} = \ham_{\tilde{\omega}} + \ham_{\tilde{J}},
\end{equation}
where the effective on-site Hamiltonian $\ham_{\tilde{\omega}}$ defined in Eq.\ (\ref{eq:boundary_term}) takes into account the boundary effects while the effective hopping Hamiltonian $\ham_{\tilde{J}}$ defined in Eq.\ (\ref{eq:effective_hopping}) gives the $N$th order coupling. Note that only the relative values of the effective on-site energies $\tilde{\omega}_\ell$ at adjacent sites are important if we study the dynamics starting from a definite Fock state $\ket{N_{\ell_0}}$. Knowing these is enough to determine the zeroth-order eigenstates, and the possible differences in the actual values of $\tilde{\omega}_\ell$ only show up in the global phase of the time-evolved state.

\subsection{Disorder}
\added{
Finally, let us briefly analyze the effect of disorder on the motion of the stack. Following the general discussion of App.\ \ref{app:theory_disorder}, we need to consider both $\hat{P}_\mathcal{A} \ham_D \hat{P}_\mathcal{A}$ and $\hat{Q}_\mathcal{A} \ham_D \hat{Q}_\mathcal{A}$. The former enters the perturbation analysis as is (through $\hat{K}_1$), while the leading term containing the latter is $\hat{P}_\mathcal{A} \ham_J \hat{W}^{(0)} \ham_D \hat{W}^{(0)} \ham_J \hat{P}_\mathcal{A}$ (through $\hat{K}_3$). Now,
\begin{align}
\hat{P}_\mathcal{A} \ham_D \hat{P}_\mathcal{A} \ket{N_\ell} &= \hbar \Delta_\ell \ket{N_\ell}, \\
\hat{P}_\mathcal{A} \ham_J \hat{W}^{(0)} \ham_D \hat{W}^{(0)} \ham_J \hat{P}_\mathcal{A} \ket{N_\ell} &= \hbar \Delta'_\ell \ket{N_\ell},
\end{align}
where}
\begin{align}
\Delta_\ell &= N \left[\delta \omega_\ell - \frac{\delta U_\ell}{2} (N - 1) \right], \\
\Delta'_\ell &= \frac{N}{N - 1} \left(\frac{J}{U} \right)^2 \bigg[\frac{\delta U_{\ell - 1} + 2 \delta U_\ell + \delta U_{\ell + 1}}{2} \nonumber \\
&\hspace{80pt}+ \frac{\Delta_{\ell - 1} + 2(N - 1) \Delta_\ell + \Delta_{\ell + 1}}{N (N - 1)} \bigg].
\end{align}
\added{
We see that if the on-site energies can be tuned locally, it is, at least in principle, possible to make $\Delta_\ell$ arbitrarily small even if we cannot control the local interactions. More precisely, $\Delta_\ell$ vanishes if we set $\delta \omega_\ell = \delta U_\ell (N - 1) / 2$. But doing so still leaves us with nonzero $\Delta'_\ell$, meaning that the disorder in $U_\ell$ cannot be completely eliminated via $\omega_\ell$.

As discussed in App.\ \ref{app:theory_disorder}, disorder begins to affect the dynamics of the stack when either $\Delta_\ell$ or $\Delta'_\ell$ start to approach the magnitude of $\tilde{J}$. Without flux tuning, $\delta \omega_\ell$ dominates $\delta U_\ell$. On the other hand, if flux tuning is available, there is still some experimental lower bound which limits how accurately $\omega_\ell$ can be adjusted. In both cases we can write $\Delta_\ell \sim N D_\omega$ if we understand $D_\omega$ to mean either the inherent manufacturing disorder in the on-site energies of the devices or the accuracy with which $\omega_\ell$ can be controlled, depending on the situation. Similarly, we have $\Delta'_\ell \sim (J / U)^2 (D_U + D_\omega)$, where $D_U$ is the disorder in the local interaction strengths due to the manufacturing process. These relations, together with the definition of $\tilde{J}$, can then be used to estimate whether disorder prevents the motion of the stack.
}

\section{Interplay between a boson stack and a single boson}\label{app:stack+boson}
\subsection{Degenerate states}
The anharmonicity of the initial state $\ket{N_{\ell_{N 0}}, 1_{\ell_{1 0}}}$ is $-N(N - 1)/2$. Clearly all the states $\ket{N_{\ell_N}, 1_{\ell_1}}$ with $\ell_N, \ell_1 = 1, \ldots, L$ and $\ell_1 \neq \ell_N$ share the same value, but are there any other Fock states $\ket{n_1 n_2 \ldots n_L}$ in this manifold when the total number of bosons is fixed? Well, the anharmonicity of the states $\ket{(N + 1)_\ell}$ is $-(N + 1)N / 2 \neq -N(N - 1) / 2$. For all the other states, we can assume that the maximum value of $n_\ell$, attained at some site $\tilde{\ell}$, is $N - m$, with $1 \leq m \leq N - 1$. First, if $m = N - 1$, every $n_\ell$ is either zero or one, and the anharmonicity vanishes. For $m < N - 1$, we obtain
\begin{align}
    \sum_{\ell = 1}^{L} n_\ell^2 &= (N - m)^2 + \sum_{\ell \neq \tilde{\ell}} n_\ell^2 \leq (N - m)^2 + (m + 1)^2 \notag \\
    &= N^2 - 2 m (N - m - 1) + 1 < N^2,
\end{align}
where we have applied the inequality (\ref{eq:occupation_squares_inequality}) on the first line [This inequality is actually stricter, since here we also have the restriction $n_\ell \leq n_{\tilde{l}}$, whereas Eq.\ (\ref{eq:occupation_squares_inequality}) was derived for unconstrained $n_\ell$. However, this milder version is enough for our purposes]. There are therefore no nontrivial states in the anharmonicity manifold $\mathcal{A}$.

\subsection{Perturbation analysis}
Let us then apply Eq.\ (\ref{eq:schrödinger_n}) inside the anharmonicity manifold $\mathcal{A} = \mathrm{span}\left\{\ket{N_{\ell_N}, 1_{\ell_1}} | \ell_N, \ell_1 = 1, \ldots, L ; \ell_1 \neq \ell_N \right\}$, and make use of the representation (\ref{eq:Hamiltonian_nth_order}) of $\ham_\mathcal{A}^{(n)}$. In the following, unless otherwise mentioned, we assume that $N > 2$.

\subsubsection{First-order analysis}\label{app:boson_quasiparticle_first_order}
At first order, we need to solve
\begin{equation}
    \hat{P}_\mathcal{A} \ham_J \hat{P}_\mathcal{A} \ket{E_\mathcal{A}^{(0)}} = E^{(1)} \ket{E_\mathcal{A}^{(0)}}.
\end{equation}
Due to the presence of the additional boson, the matrix $\hat{P}_\mathcal{A} \ham_J \hat{P}_\mathcal{A}$ does not vanish identically as it did when there was only the quasiparticle present, and there are thus first-order corrections to the energy.

We see that $\hat{P}_\mathcal{A} \ham_J \hat{P}_\mathcal{A}$ has a block diagonal structure: for a fixed value of $\ell_N$, the states with $\ell_1 < \ell_N$ form a coupled block, as do the states with $\ell_1 > \ell_N$. There is no coupling between the blocks since it would require more than one application of the hopping Hamiltonian. For example, if $L = 4$, we have six separate blocks with no coupling between them: $\left\{\ket{N 1 0 0}, \ket{N 0 1 0}, \ket{N 0 0 1} \right\}$, $\left\{\ket{1 N 0 0} \right\}$, $\left\{\ket{0 N 1 0}, \ket{0 N 0 1} \right\}$, $\left\{\ket{1 0 N 0}, \ket{0 1 N 0} \right\}$, $\left\{\ket{0 0 N 1} \right\}$, and $\left\{\ket{1 0 0 N}, \ket{0 1 0 N}, \ket{0 0 1 N} \right\}$.

Within each such block, the problem simply reduces to a one-boson problem, but the effective length of the chain is now determined by the location of the quasiparticle. Continuing the example with $L = 4$, the matrix elements of $\hat{P}_\mathcal{A} \ham_J \hat{P}_\mathcal{A}$ within $\left\{\ket{N 1 0 0}, \ket{N 0 1 0}, \ket{N 0 0 1} \right\}$ are exactly the same as for a single particle in a chain of length three. The quasiparticle thus acts as an effective boundary which the boson cannot cross.

For a fixed value of $\ell_N$, the effective length of the chain is $\ell_N - 1$ if the boson is to the left of the quasiparticle and $L - \ell_N$ if the boson is to the right. Using Eq.\ (\ref{eq:single-particle_energies}), the first-order energies are thus
\begin{align}
    E^{(1)}_{-, k_-, \ell_N} / \hbar &= 2 J \cos\left(\frac{\pi k_-}{\ell_N} \right), \label{eq:first_order_energy_-} \\
    E^{(1)}_{+, k_+, \ell_N} / \hbar &= 2 J \cos\left(\frac{\pi k_+}{L - \ell_N + 1} \right), \label{eq:first_order_energy_+}
\end{align}
where $k_- = 1, \ldots, \ell_N - 1$; $k_+ = 1, \ldots, L - \ell_N$; and $\ell_N = 1, \ldots, L$. The subscript $-$ ($+$) refers to the case where the boson is to the left (right) of the quasiparticle. The corresponding eigenstates are
\begin{align}
    \ket{\varepsilon_{-, k_-, \ell_N}} &= \sqrt{\frac{2}{\ell_N}} \sum_{\ell_1 = 1}^{\ell_N - 1} \sin\left(\frac{\pi \ell_1 k_-}{\ell_N} \right) \ket{N_{\ell_N}, 1_{\ell_1}}, \label{eq:first_order_state_-} \\
    \ket{\varepsilon_{+, k_+, \ell_N}} &= \sqrt{\frac{2}{L - \ell_N + 1}} \sum_{\ell_1 = \ell_N + 1}^{L} \sin\left[\frac{\pi (\ell_1 - \ell_N) k_+}{L - \ell_N + 1} \right] \nonumber \\
    &\hspace{100pt} \times \ket{N_{\ell_N}, 1_{\ell_1}}. \label{eq:first_order_state_+}
\end{align}
Remember, however, that these are not necessarily the true zeroth-order states of our full problem. Just like we initially grouped the Fock states into different anharmonicity manifolds according to their zeroth-order energies, here we need to group the states $\ket{\varepsilon_{\pm, k_\pm, \ell_N}}$ according to their first-order energies. All we know at this point is that the zeroth-order eigenstates $\ket{E^{(0)}}$ are some linear combinations of the states belonging to the same eigenspace of the first-order Hamiltonian $\hat{P}_\mathcal{A} \ham_J \hat{P}_\mathcal{A}$, and the weights of these superpositions are determined at higher orders. In our example of $L = 4$, there are, for example, four states with zero first-order energy: $(\ell_N = 1, k_+ = 2)$, $(\ell_N = 2, k_- = 1)$, $(\ell_N = 3, k_+ = 1)$, and $(\ell_N = 4, k_- = 2)$. There are thus four zeroth-order states which have vanishing first-order energy, each some linear combination of the states (\ref{eq:first_order_state_-}) and (\ref{eq:first_order_state_+}) with the above parameters.

Let us then consider the initial state $\ket{N_{\ell_{N 0}}, 1_{\ell_{1 0}}}$. Due to symmetry, no generality is lost if we assume $\ell_{1 0} < \ell_{N 0}$. The initial state now belongs to the same block with $\ell_{N 0} - 2$ other states $\ket{N_{\ell_{N 0}}, 1_{\ell_1}}$ with $\ell_1 < \ell_{N 0}$, and is thus involved in $\ell_{N 0} - 1$ eigenstates $\ket{\varepsilon_{-, k_-, \ell_{N 0}}}$ of Eq.\ (\ref{eq:first_order_state_-}). More precisely, we have
\begin{equation}\label{eq:boson_quasiparticle_initial_state}
    \ket{N_{\ell_{N 0}}, 1_{\ell_{1 0}}} = \sqrt{\frac{2}{\ell_{N 0}}} \sum_{k_- = 1}^{\ell_{N 0} - 1} \sin\left(\frac{\pi \ell_{1 0} k_-}{\ell_{N 0}} \right) \ket{\varepsilon_{-, k_-, \ell_{N 0}}}.
\end{equation}
The final part of the first-order analysis is to find the eigenspaces these states belong to, that is, find all the states which have energies $2 \hbar J \cos(\pi k_- / \ell_{N 0})$, $k_- = 1, \ldots, \ell_{N 0} - 1$, respectively. This is not a trivial task since it involves solving rational equations of the form $k / L = k' / L'$. What we can do, however, is to first rule out some of the states and proceed from there.

Specifically, let us show that the blocks where the position of the quasiparticle differs by one ($\ell_N$ vs.\ $\ell_N \pm 1$) while the boson stays on the same side of the quasiparticle always belong to different eigenspaces. In our example of $L = 4$ this means that the blocks $\left\{\ket{N 1 0 0}, \ket{N 0 1 0}, \ket{N 0 0 1} \right\}$ and $\left\{\ket{0 0 N 1} \right\}$ share no common first-order energies with the block $\left\{\ket{0 N 1 0}, \ket{0 N 0 1} \right\}$. Let us here prove this fact when the boson is to the left of the quasiparticle. The other case can be proved with exactly the same logic. Now,
\begin{align*}
    &\phantom{\Leftrightarrow} & 2 J \cos\left(\frac{\pi k_-}{\ell_N} \right) &= 2 J \cos\left(\frac{\pi k'_-}{\ell_N \pm 1} \right) \\
    &\Leftrightarrow & \frac{k_-}{\ell_N} &= \frac{k'_-}{\ell_N \pm 1} \\
    &\Leftrightarrow & k_- &= \pm (k'_- - k_-) \ell_N,
\end{align*}
implying that $k_-$ has to be a multiple of $\ell_N$. But this is impossible since $k_- = 1, \ldots, \ell_N - 1$, proving our claim. It turns out that this small observation gets us really far, as we shall see in the following.

\begin{figure}
    \centering
    \includegraphics[width=\columnwidth]{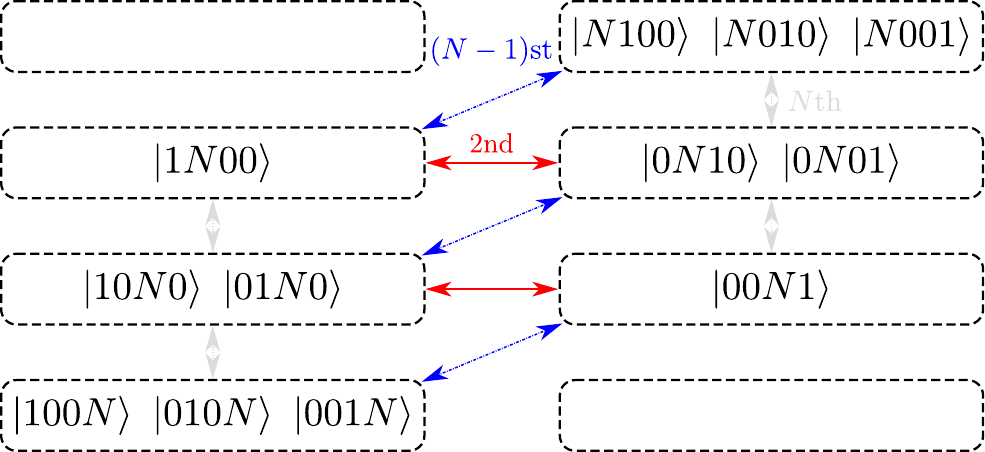}
    \caption{A schematic representation of the block structure of the states at first order in a chain with a quasiparticle and a boson, here presented for $L = 4$. In the left (right) column, the boson is located to the left (right) of the quasiparticle. The arrows represent possible couplings between blocks at orders less than or equal to $N + 2$. As discussed in the text, the vertical gray arrows are never active since the corresponding blocks do not share any first-order energies. Moreover, depending on the particular first-order eigenspace under consideration, either the red or the blue arrows are inactive due to the same reason. This means that up to $(N + 2)$nd order, the zeroth-order eigenstates are always a mixture of states from exactly two blocks.}
    \label{fig:boson_quasiparticle_couplings}
\end{figure}

The different blocks of states become coupled at different orders, providing possible channels for mixing of the states within a given first-order eigenspace. But we only need to worry about the couplings which are of equal or less order than where we want to eventually proceed to in our analysis. Since the motion of the quasiparticle in isolation would occur at order $N$, it is also the natural order up to which we want to carry our analysis here. 
Neglecting couplings weaker than $(N + 2)$nd order, we end up with the coupling diagram of Fig.\ \ref{fig:boson_quasiparticle_couplings}. The vertical $N$th-order couplings between neighboring blocks (light gray arrows), even though in principle there, are never actually active since, as we saw above, the blocks share no states with common first-order energies and there are thus no states within one eigenspace to couple to. Due to exactly the same reason, either the horizontal second-order couplings (red arrows) or the diagonal $(N - 1)$st order couplings (blue arrows) are always inactive, depending on the eigenspace considered.

To see this, arrange the blocks in an array like in Fig.\ \ref{fig:boson_quasiparticle_couplings} and denote by $b_{i j}$ the block on the $i$th row and $j$th column, with $i = 1, \ldots, L$ and $j = 1, 2$. Since the initial state belongs to the left column, pick any block $b_{i 1}$ and choose one of the first-order states (\ref{eq:first_order_state_-}) from there. Now, if the horizontally adjacent block $b_{i 2}$ has a state with matching energy, then we know from the earlier analysis that the blocks $b_{(i \pm 1) 2}$ cannot have such states, and therefore the row $i$ is isolated from all the other blocks. Similarly, if the block $b_{(i - 1) 2}$ has a state with the same energy, the block $b_{i 2}$ does not, and therefore the pair of blocks $b_{i j}$, $b_{(i - 1) 2}$ is isolated from every other block.

Finally, we note that for fixed $\ell_N$, the energies (\ref{eq:first_order_energy_-}) are non-degenerate, as are the energies (\ref{eq:first_order_energy_+}). In other words, each block has no two states with equal energies, and the higher-order analysis is thus reduced to a number of two-state problems.

To sum up, the first-order analysis tells us that the number of zeroth-order eigenstates of the full Hamiltonian contributing to the initial state $\ket{N_{\ell_{N 0}}, 1_{\ell_{1 0}}}$ with $\ell_{1 0} < \ell_{N 0}$ is between $\ell_{N 0} - 1$ and $2(\ell_{N 0} - 1)$. The states fall into three categories, one for each $k_- = 1, \ldots, l_{N 0} - 1$: (i) If $k_+^{(\mathrm{i})} \equiv k_- (L - \ell_{N 0} + 1) / \ell_{N 0} \in \{1, \ldots, L - \ell_{N 0} \}$, there are two eigenstates which are linear combinations of the states $\ket{\varepsilon_{-, k_-, \ell_{N 0}}}$ and $\ket{\varepsilon_{+, k_+^{(\mathrm{i})}, \ell_{N 0}}}$. (ii) If $k_+^{(\mathrm{ii})} \equiv k_- (L - \ell_{N 0} + 2) / \ell_{N 0} \in \{1, \ldots, L - \ell_{N 0} + 1 \}$, there are two eigenstates which are linear combinations of the states $\ket{\varepsilon_{-, k_-, \ell_{N 0}}}$ and $\ket{\varepsilon_{+, k_+^{(\mathrm{ii})}, \ell_{N 0} - 1}}$. (iii) If neither of the conditions hold, there is a single eigenstate given simply by $\ket{\varepsilon_{-, k_-, \ell_{N 0}}}$.

For states in categories (i) and (iii), the quasiparticle always stays in the initial position $\ell_{N 0}$, but for states in category (ii), it is possible that the quasiparticle moves to $\ell_{N 0} - 1$.

\subsubsection{States of category (ii)}
Going forward in perturbation theory, let us first consider the states belonging to category (ii), since this turns out to be a special case. To this end, let us assume that $k_+^{(\mathrm{ii})} \equiv k_- (L - \ell_{N 0} + 2) / \ell_{N 0} \in \{1, \ldots, L - \ell_{N 0} + 1 \}$ for some $k_- = 1, \ldots, \ell_{N 0} - 1$. As we saw above, we can restrict our analysis to the two-dimensional manifold spanned by the states $\ket{\varepsilon_{-, k_-, \ell_{N 0}}}$ and $\ket{\varepsilon_{+, k_+^{(\mathrm{ii})}, \ell_{N 0} - 1}}$. The states couple at $(N - 1)$st order, so in order to mix, the degeneracy has to stay intact up to that order. In other words, if these two states have different energies at any order below $(N - 1)$st, they are not entangled, and are both individually zeroth-order eigenstates. But intuitively it seems clear that the degeneracy can only be maintained if there is enough symmetry present, that is, the effective lengths of the chains $\ell_{N 0} - 1$ and $L - \ell_{N 0} + 1$ and the wavenumbers $k_-$ and $k_+^{(\mathrm{ii})}$ have to be equal, respectively. In this case, possible when $\ell_{N 0} = L/2 + 1$, the states are identical at every order from the point of view of perturbation theory. The case $N = 3$ is of course exceptional since the coupling occurs already at second order. Let us check that our intuition is indeed correct.

A direct calculation shows that the diagonal elements of the $2 \times 2$ matrix $H_\mathcal{A}^{(2)}$ are given by
\begin{widetext}
\begin{align}
    \braket{\varepsilon_{-, k_-, \ell_N} | \ham_\mathcal{A}^{(2)} | \varepsilon_{-, k_-, \ell_N}} &= - \hbar U \left(\frac{J}{U} \right)^2 \left\{\frac{(2 - \delta_{\ell_N, L}) N}{N - 1} - \frac{2}{\ell_N} \sin^2 \left(\frac{\pi k_-}{\ell_N} \right) \left(\frac{2 N^2 - 1}{N(N - 1)} - \frac{2 N}{N - 2} \right) \right\}, \label{eq:boson_quasiparticle_2nd_order_diagonal_-} \\
    \braket{\varepsilon_{+, k_+, \ell_N} | \ham_\mathcal{A}^{(2)} | \varepsilon_{+, k_+, \ell_N}} &= \braket{\varepsilon_{-, k_+, L - \ell_N + 1} | \ham_\mathcal{A}^{(2)} | \varepsilon_{-, k_+, L - \ell_N + 1}}. \label{eq:boson_quasiparticle_2nd_order_diagonal_+}
\end{align}
\end{widetext}
Using these, we can now find out the condition for the degeneracy to remain intact by setting $\braket{\varepsilon_{-, k_-, \ell_{N 0}} | \ham_\mathcal{A}^{(2)} | \varepsilon_{-, k_-, \ell_{N 0}}}$ equal to $\braket{\varepsilon_{+, k_+^{(\mathrm{ii})}, \ell_{N 0} - 1} | \ham_\mathcal{A}^{(2)} | \varepsilon_{+, k_+^{(\mathrm{ii})}, \ell_{N 0} - 1}}$. A straightforward analysis reveals that this can only be true if $\ell_{N 0} = L/2 + 1$, proving our earlier intuition right.

Let us therefore assume that $\ell_{N 0} = L/2 + 1$ and $N > 3$. In this case, we actually see that all the relevant zeroth-order eigenstates fall into category (ii) since we have $k_+^{(\mathrm{ii})} = k_-$. No orders less than $N - 1$ provide us any further information regarding the eigenstates since the degeneracy between $\ket{\varepsilon_{-, k_-, L/2 + 1}}$ and $\ket{\varepsilon_{+, k_-, L/2}}$ is not lifted due to symmetry.

At order $N - 1$, we finally obtain coupling between the states. Clearly the matrix $H_\mathcal{A}^{(N - 1)}$ is of the form $\begin{pmatrix}a & b \\ b & a\end{pmatrix}$, where $a$ and $b$ are real. This immediately tells us that the zeroth-order eigenstates are given by
\begin{equation}
    \ket{E^{(0)}_{\pm, k}} = \frac{\ket{\varepsilon_{-, k, L/2 + 1}} \pm \ket{\varepsilon_{+, k, L/2}}}{\sqrt{2}}
\end{equation}
for $k = 1, \ldots, L$, and so any initial state $\ket{\varepsilon_{-, k, L/2 + 1}}$ will oscillate between itself and $\ket{\varepsilon_{+, k, L/2}}$ at angular frequency
\begin{equation}
\begin{split}
    \Omega_k &\equiv \frac{2}{\hbar} \braket{\varepsilon_{+, k, L/2} | \ham_\mathcal{A}^{(N - 1)} | \varepsilon_{-, k, L/2 + 1}} \\
    &= \frac{4 (-1)^{k + 1}}{L/2 + 1} \sin^2\left(\frac{\pi k}{L/2 + 1} \right) \Xi,
\end{split}
\end{equation}
where
\begin{align}
    \Xi &\equiv \braket{N_{L/2}, 1_{L/2 + 1} | \ham_\mathcal{A}^{(N - 1)} / \hbar | 1_{L/2}, N_{L/2 + 1}} \notag \\
    &= \bra{N_{L/2}, 1_{L/2 + 1} } J \hat{a}_{L/2}^\dagger \hat{a}_{L/2 + 1} \notag\\
    &\hspace{40pt}\times \left(\hat{W}^{(0)} \hbar J \hat{a}_{L/2}^\dagger \hat{a}_{L/2 + 1} \right)^{N - 2} \ket{1_{L/2}, N_{L/2 + 1}} \notag \\
    &= J (\hbar J)^{N - 2} \prod_{m = 0}^{N - 2} \sqrt{N - m} \sqrt{m + 2} \notag\\
    &\hspace{47pt} \times \prod_{m = 1}^{N - 2} [-\hbar U m (N - m - 1)]^{-1} \notag\\
    &= (-1)^N U \frac{N (N - 1)}{(N - 2)!} \left(\frac{J}{U} \right)^{N - 1}
\end{align}
is the coupling constant between the Fock states states $\ket{1_{L/2}, N_{L/2 + 1}}$ and $\ket{N_{L/2}, 1_{L/2 + 1}}$. 

Since the actual initial state $\ket{1_{\ell_{1 0}}, N_{L/2 + 1}}$ is a superposition of the states $\ket{\varepsilon_{-, k, L/2 + 1}}$ with different $k$ [see Eq.\ (\ref{eq:boson_quasiparticle_initial_state})], the dynamics is a bit more involved than the above due to the relative phase differences caused by $E^{(1)}, \ldots, E^{(N - 2)}$. Namely,
\begin{align}
    \ket{\Psi(t)} &\approx \sqrt{\frac{2}{L/2 + 1}} \sum_{k = 1}^{L/2} \sin\left(\frac{\pi \ell_{1 0} k}{L/2 + 1} \right) \notag \\
    &\hspace{20pt}\times e^{-i [E^{(0)} + E_k^{(1)} + \ldots + E_k^{(N - 2)}] t / \hbar} \\
    &\hspace{20pt}\times \frac{e^{-i E_{+, k}^{(N - 1)} t / \hbar} \ket{E^{(0)}_{+, k}} + e^{-i E_{-, k}^{(N - 1)} t / \hbar} \ket{E^{(0)}_{-, k}}}{\sqrt{2}}, \notag
\end{align}
where $E_k^{(n)}$ for $n = 1, \ldots, N - 2$ are the common $n$th-order energies of $\ket{E^{(0)}_{\pm, k}}$. Now, unlike in the case of the sole quasiparticle, these intermediate energies depend on the summation index, and we cannot simply factor out an insignificant phase factor.

We have already calculated $E_k^{(1)}$. The rest are formally simple,
\begin{equation}
    E_k^{(n)} = \braket{\varepsilon_{-, k, L/2 + 1} | \ham_\mathcal{A}^{(n)} | \varepsilon_{-, k, L/2 + 1}}.
\end{equation}
If the knowledge of the full state is important, one can calculate these either numerically, symbolically, or by hand if $N$ is small enough. We shall not study this any further here, the important point is that on the time scale of $2 \pi / \Xi$, the quasiparticle can move between the sites $L/2 + 1$ and $L/2$ while the boson performs much more rapid motion between the quasiparticle and the boundary.

The fact that we have not calculated all the intermediate energies also means that our knowledge of the effective Hamiltonian is incomplete. The hopping Hamiltonian $\ham_J$ defined in Eq.\ (\ref{eq:ham}) and the nearest-neighbor interaction $\ham_V$ defined in Eq.\ (\ref{eq:nearest_neighbor_interaction}) correctly reproduce the first-order and second order results, respectively. Similarly, the exchange Hamiltonian $\ham_\Xi$ defined in Eq.\ (\ref{eq:exchange_stack+boson}) correctly couples the states at $(N - 1)$st order. The sum of these three therefore gives us the true zeroth-order states, but the energies are incorrect. More generally, we must write
\begin{equation}
\ham_{\mathrm{eff}} = \ham_J + \ham_V + \ham_\Xi + \ham_{\mathrm{eff}}'.
\end{equation}
Here $\ham_{\mathrm{eff}}'$ contains terms of order $3, \ldots, N - 1$, for example longer-range hopping terms for the boson. If $N$ is not too large, these can be written down analytically or symbolically, but for numerical analysis it is easiest to just use the definition (\ref{eq:eff_ham}) since we already know the states.

\subsubsection{States of category (i) or (iii)}
Let us then assume $\ell_{N 0} \neq L/2 + 1$ and $N > 3$. We saw above that in this case, we do not have to worry about the states of category (ii) since the states $\ket{\varepsilon_{-, k_-, \ell_{N 0}}}, \ket{\varepsilon_{+, k_+^{(\mathrm{ii})}, \ell_{N 0} - 1}}$ never mix. The only nontrivial possibility is therefore category (i).

Let us assume that $k_+^{(\mathrm{i})} \equiv k_- (L - \ell_{N 0} + 1) / \ell_{N 0} \in \{1, \ldots, L - \ell_{N 0} \}$ for some $k_- = 1, \ldots, \ell_{N 0} - 1$. Note that we must have $\ell_{N 0} \neq 1, L$. In this case, we know that there are two eigenstates which are linear combinations of the states $\ket{\varepsilon_{-, k_-, \ell_{N 0}}}$ and $\ket{\varepsilon_{+, k_+^{(\mathrm{i})}, \ell_{N 0}}}$. Furthermore, since these states are only two hops away from each other, the eigenstates can be solved already at second order.

The off-diagonal term
\begin{align}
    &\braket{\varepsilon_{+, k_+^{(\mathrm{i})}, \ell_{N 0}} | \ham_\mathcal{A}^{(2)} | \varepsilon_{-, k_-, \ell_{N 0}}} = \frac{2 (-1)^{k_- + 1}}{\sqrt{\ell_{N 0} (L - \ell_{N 0} + 1)}} \notag \\
    &\hspace{120pt} \times \sin^2\left(\frac{\pi k_-}{\ell_{N 0}} \right) \hbar \, T \label{eq:boson_quasiparticle_2nd_order_diagonal_tunneling}
\end{align}
of the matrix $H_\mathcal{A}^{(2)}$ is present due to the nonzero coupling term
\begin{align}
    T &\equiv \braket{N_{\ell_{N 0}}, 1_{\ell_{N 0} + 1} | \ham_\mathcal{A}^{(2)} / \hbar | 1_{\ell_{N 0} - 1}, N_{\ell_{N 0}}} \notag \\
    &= \bra{N_{\ell_{N 0}}, 1_{\ell_{N 0} + 1} } \Big( J \hat{a}_{\ell_{N 0} + 1}^\dagger \hat{a}_{\ell_{N 0}} \hat{W}^{(0)} \hbar J \hat{a}_{\ell_{N 0}}^\dagger \hat{a}_{\ell_{N 0} - 1} \notag\\
    &\hspace{20pt} + J \hat{a}_{\ell_{N 0}}^\dagger \hat{a}_{\ell_{N 0} - 1} \hat{W}^{(0)} \hbar J \hat{a}_{\ell_{N 0} + 1}^\dagger \hat{a}_{\ell_{N 0}} \Big) \ket{1_{\ell_{N 0} - 1}, N_{\ell_{N 0}}} \notag\\
    &= - \frac{U}{N (N - 1)} \left(\frac{J}{U} \right)^2
\end{align}
between the Fock states $\ket{1_{\ell_{N 0} - 1}, N_{\ell_{N 0}}}$ and $\ket{N_{\ell_{N 0}}, 1_{\ell_{N 0} + 1}}$, describing tunneling of the boson through the quasiparticle.

How strong is the mixing between the states $\ket{\varepsilon_{-, k_-, \ell_{N 0}}}$ and $\ket{\varepsilon_{+, k_+^{(\mathrm{i})}, \ell_{N 0}}}$? The familiar Rabi formula tells us that starting from the former, the maximum probability of observing the system in the latter is $P_{- \to +} = 1 / \{1 + [(H_{- -} - H_{+ +}) / 2 |H_{+ -}|]^2\}$, where we have denoted the matrix elements of $\ham_\mathcal{A}^{(2)}$ by $H_{+ -} = \braket{\varepsilon_{+, k_+^{(\mathrm{i})}, \ell_{N 0}} | \ham_\mathcal{A}^{(2)} | \varepsilon_{-, k_-, \ell_{N 0}}}$ and so on and so forth. Using Eqs.\ (\ref{eq:boson_quasiparticle_2nd_order_diagonal_-}), (\ref{eq:boson_quasiparticle_2nd_order_diagonal_+}), and (\ref{eq:boson_quasiparticle_2nd_order_diagonal_tunneling}), we obtain
\begin{equation}
    P_{- \to +} = \frac{1}{1 + \left(\sqrt{\frac{\ell_{N 0}}{L - \ell_{N 0} + 1}} - \sqrt{\frac{L - \ell_{N 0} + 1}{\ell_{N 0}}} \right)^2 \left(\frac{N^2}{N - 2} + \frac{1}{2} \right)^2}.
\end{equation}
We see that the denominator is large in most cases. The only exception is when the square root terms cancel each other out, leaving $P_{- \to +} = 1$. This happens when $\ell_{N 0} = (L + 1) / 2$, that is, when the quasiparticle sits right in the middle of the chain. 

Now, the initial state $\ket{N_{\ell_{N 0}}, 1_{\ell_{1 0}}}$ is a linear combination of the states $\ket{\varepsilon_{-, k_-, \ell_{N 0}}}$ as shown by Eq.\ (\ref{eq:boson_quasiparticle_initial_state}). If $\ell_{N 0} = (L + 1) / 2$, all the eigenstates belong to category (i) and we expect the tunneling to be significant. In other cases, some of the states are of category (iii), contributing nothing to the tunneling, and even for the states in category (i) the effect is small. We can thus, to a good approximation, ignore the tunneling if $\ell_{N 0} \neq (L + 1) / 2$.

The effective Hamiltonian recreating the above results is given by
\begin{equation}
    \ham_{\mathrm{eff}} = \ham_J + \ham_V + \ham_T,
\end{equation}
where the hopping Hamiltonian $\ham_J$ defined in Eq.\ (\ref{eq:ham}) again gives the correct first-order behavior, while the nearest-neighbor interaction term $\ham_V$ and the tunneling Hamiltonian $\ham_T$, defined in Eqs.\ (\ref{eq:nearest_neighbor_interaction}) and (\ref{eq:tunneling_term}), respectively, take care of the second-order analysis.

As a final note, if $N = 2$ or $N = 3$, we just need to include the exchange Hamiltonian into the effective Hamiltonian above.

\subsection{Disorder}
\added{
Let us briefly analyze the effect of disorder on pure tunneling and pure exchange. The special cases where both are present can be handled similarly.

First, because tunneling is a second-order effect, we need to only consider the operator $\hat{P}_\mathcal{A} \ham_D \hat{P}_\mathcal{A}$, which now satisfies
\begin{equation}
\begin{split}
\hat{P}_\mathcal{A} \ham_D \hat{P}_\mathcal{A} \ket{N_{\ell_N}, 1_{\ell_1}} &= \bigg[\delta \omega_{\ell_N} N - \frac{\delta U_{\ell_N}}{2} N (N - 1) \\
&\hspace{20pt} + \delta \omega_{\ell_1} \bigg] \ket{N_{\ell_N}, 1_{\ell_1}}.
\end{split}
\end{equation}
In general, it is no longer possible to make $\hat{P}_\mathcal{A} \ham_D \hat{P}_\mathcal{A}$ vanish identically in the whole of $\mathcal{A}$ just by adjusting $\omega_\ell$. This is because the number of bosons at any given site can be either $N$, one, or zero. In the current case, however, the stack stays put and the situation is basically the same as in the case of a single stack. We therefore see that the disorder starts to affect the tunneling dynamics when $D_\omega \sim T$.

A straightforward calculation shows that $\hat{P}_\mathcal{A} \ham_D \hat{P}_\mathcal{A}$ can be made arbitrarily small also in the subspace of $\mathcal{A}$ relevant to the case of exchange by choosing
\begin{equation}
\delta \omega_m = 
\begin{cases}
\frac{N}{2 (N + 1)} \left(\delta U_{L/2} N - \delta U_{L/2 + 1} \right) & \mathrm{if} \; m \leq L/2, \\
\frac{N}{2 (N + 1)} \left(\delta U_{L/2 + 1} N - \delta U_{L/2} \right) & \mathrm{if} \; m \geq L/2 + 1,
\end{cases}
\end{equation}
but now we cannot neglect $\hat{Q}_\mathcal{A} \ham_D \hat{Q}_\mathcal{A}$ which again first appears as $\hat{P}_\mathcal{A} \ham_J \hat{W}^{(0)} \ham_D \hat{W}^{(0)} \ham_J \hat{P}_\mathcal{A}$. Based on the analysis of the single stack, we therefore expect the disorder to start to affect the exchange dynamics when the larger of $D_\omega$ and $(J/U)^2 D_U$ becomes comparable to $\Xi$.
}

\section{Interacting boson stacks}\label{app:two_stacks}
\subsection{Degenerate states}
The anharmonicity of the initial state $\ket{N_{\ell_{N 0}}, M_{\ell_{M 0}}}$ is $-N(N - 1)/2 - M(M - 1)/2$. Unlike in the two previous cases we have considered, the anharmonicity manifold $\mathcal{A}$ does not necessarily contain only the trivial states $\ket{N_{\ell_N}, M_{\ell_M}}$. For example, in the case of $N = M = 3$, the states $\ket{4 1 1 0 \dots 0}$ etc.\ share the same anharmonicity of $-6$ as the trivial states.

The general problem of finding all the states with a given anharmonicity seems to be quite hard, since it involves solving a constrained multivariate equation among the positive integers. There are some facts, however, which can be shown quite easily. Let $\ket{n_1 n_2 \ldots n_L}$ be a nontrivial Fock state with $\sum_{\ell = 1}^{L} n_{\ell} = N + M$ and $\sum_{\ell = 1}^{L} n_{\ell}^2 = N^2 + M^2$. Without loss of generality, let us also assume $M \leq N$.

First of all, at least three sites must have nonzero occupation. If only one site was occupied with $N + M$ bosons, we would have $(N + M)^2 = N^2 + M^2$, implying the impossibility $M N = 0$. If two sites were occupied with $\tilde{N}$ and $N + M - \tilde{N}$ bosons, respectively, we would have $\tilde{N}^2 + (N + M - \tilde{N})^2 = N^2 + M^2$. But solving this quadratic equation yields either $\tilde{N} = M$ or $\tilde{N} = N$, which are the trivial states.

Next, we see that none of the sites can have $M$ or $N$ bosons. To see this, assume the contrary, say, $n_{\ell_0} = N$ for some $\ell_0$ (the case of $n_{\ell_0} = M$ follows exactly the same logic). Then $\sum_{\ell \neq \ell_0} n_{\ell}^2  = M^2$. But applying Eq.\ (\ref{eq:occupation_squares_inequality}), we must have $\sum_{\ell \neq \ell_0} n_{\ell}^2 < M^2$ (at least two of the remaining $n_{\ell}$s are nonzero, hence the strict inequality), leading to the contradiction $M < M$.

Let us then order the sites according to their occupation numbers, so that $n_{\tilde{l}_1} \geq n_{\tilde{\ell}_2} \geq \ldots \geq n_{\tilde{\ell}_L}$. Then $\sum_{\ell = 1}^{L} n_{\ell}^2 = n_{\tilde{\ell}_1}^2 + \sum_{\ell \neq \tilde{\ell}_1} n_{\ell}^2 = N^2 + M^2$. Applying again Eq.\ (\ref{eq:occupation_squares_inequality}), we have $\sum_{\ell \neq \tilde{\ell}_1} n_{\ell}^2 < (N + M - n_{\tilde{\ell}_1})^2$, and thus $N^2 + M^2 - n_{\tilde{\ell}_1}^2 < (N + M - n_{\tilde{\ell}_1})^2$. This is satisfied if either $n_{\tilde{\ell}_1} < M$ or $n_{\tilde{\ell}_1} > N$. The former is not a proper solution. If it were, we would have $N^2 + M^2 = \sum_{\ell} n_{\ell}^2 \leq \sum_{\ell} n_{\ell} n_{\tilde{\ell}_1} = n_{\tilde{\ell}_1} (N + M) < M (N + M)$, which is clearly impossible since $M \leq N$. We therefore find $n_{\tilde{\ell}_1} \geq N + 1$, that is, maximum occupation is at least $N + 1$.

We can also bound the next-highest occupation $n_{\tilde{\ell}_2}$ from above. Since $\sum_{\ell} n_{\ell}^2 = N^2 + M^2$, we have $\sum_{\ell \neq \tilde{\ell}_1} n_{\ell}^2 = N^2 + M^2 - n_{\tilde{\ell}_1}^2$. But now $\sum_{\ell \neq \tilde{\ell}_1} n_{\ell}^2 \geq n_{\tilde{\ell}_2}^2 + 1$, while $N^2 + M^2 - n_{\tilde{\ell}_1}^2 \leq N^2 + M^2 - (N + 1)^2 \leq M^2 - 2 M - 1$. We therefore must have $n_{\tilde{\ell}_2}^2 + 1 \leq M^2 - 2 M - 1$, and so $n_{\tilde{\ell}_2}^2 \leq (M - 1)^2 - 3$. Hence, we obtain the bound $n_{\tilde{\ell}_2} \leq M - 2$, that is, the second-highest occupation is at most $M - 2$.

To recap, we have shown that a nontrivial Fock state $\ket{n_1 n_2 \ldots n_L}$ with total boson number $N + M$ and anharmonicity $-N(N - 1)/2 - M(M - 1)/2$ must have at least three occupied sites, the maximum occupation is at least $N + 1$, and the second-highest occupation is not greater than $M - 2$. If we find one such set of $n_\ell$, then of course all the different permutations among the sites are also valid solutions.

The crucial question here is at what order do the nontrivial states couple to the trivial ones. If this coupling occurs at order higher than two, then the degeneracy between the trivial and nontrivial states is almost certainly broken at second order (compare with the analysis of second order energies in the case of $M = 1$ in App.~\ref{app:stack+boson}) and we do not need to worry about the nontrivial states at all.

Coupling at first order is not possible since we need to move at least one boson from the site $\tilde{\ell}_1$ and completely clear at least the site $\tilde{\ell}_3$. Similarly, second-order coupling is not possible if more than three sites are occupied. 

If three sites are occupied, second-order coupling is possible if $n_{\tilde{\ell}_1} = N + 1$ and $n_{\tilde{\ell}_3} = 1$. This leaves $n_{\tilde{\ell}_2} = M - 2$, and the site $\tilde{\ell}_2$ must lie between the sites $\tilde{\ell}_1$ and $\tilde{\ell}_3$. We must, of course, satisfy $(N + 1)^2 + (M - 2)^2 + 1^2 = N^2 + M^2$, and therefore $M = (N + 3)/2$.

In order to get a rough estimate for the amount of mixing between the trivial and nontrivial states, let us consider the pair of states $\ket{N_\ell, M_{\ell + 1}}$, $\ket{(N + 1)_\ell, (M - 2)_{\ell + 1}, 1_{\ell + 2}}$ at second order, with $M = (N + 3)/2$. The nontrivial state here is of course coupled to other nontrivial states already at first order and we should thus consider one of the eigenstates, but this simplified analysis should get us in the ballpark. Just like in the case of the quasiparticle and the boson, we use the Rabi formula to calculate the maximum probability of being in the state $\ket{(N + 1)_\ell, (M - 2)_{\ell + 1}, 1_{\ell + 2}}$ if we start from the state $\ket{N_\ell, M_{\ell + 1}}$. Assuming none of the sites is at the boundary, we obtain

\begin{widetext}
\begin{align}
    \braket{N_\ell, M_{\ell + 1} | \ham_\mathcal{A}^{(2)} | N_\ell, M_{\ell + 1}} &= - \hbar U \left(\frac{J}{U} \right)^2 \left[\frac{N (N + 5)}{N - 5} - \frac{N^2 + 3 N + 3}{N - 1} + \frac{N + 3}{N + 1} \right], \\
    \braket{(N \! \! + \! \! 1)_\ell, (M \! \! - \! \! 2)_{\ell + 1}, 1_{\ell + 2} | \ham_\mathcal{A}^{(2)} | (N \! \! + \! \! 1)_\ell, (M \! \! - \! \! 2)_{\ell + 1}, 1_{\ell + 2}} &= - \hbar U \left(\frac{J}{U} \right)^2 \left[\frac{N + 1}{N} + \frac{5 N + 7}{N + 5} + \frac{2 (N - 1)}{N - 5} - \frac{N + 1}{N - 1} \right], \\
    \braket{(N \! \! + \! \! 1)_\ell, (M \! \! - \! \! 2)_{\ell + 1}, 1_{\ell + 2} | \ham_\mathcal{A}^{(2)} | N_\ell, M_{\ell + 1}} &= - \hbar U \left(\frac{J}{U} \right)^2 \left[ \frac{-2 \sqrt{N + 3}}{N - 1} \right].
\end{align}
\end{widetext}
Here we have assumed $N \neq M + 1$. In the case $N = M + 1$ (and thus $N = 5$, $M = 4$), the degeneracy is broken already at first order since $\ket{N_\ell, (N - 1)_{\ell + 1}}$ and $\ket{(N - 1)_\ell, N_{\ell + 1}}$ are coupled, producing first-order energies $\pm \hbar J N$ which are always greater in magnitude than the first-order energies of the nontrivial states. The latter must lie between $\pm 3 \hbar J$ as can be seen, for example, by using Weyl's inequality \cite{parlett1980}.

Using these matrix elements and the Rabi formula, we see that the maximum probability $P_{\mathrm{nontriv}}$ of being in the nontrivial state $\ket{(N + 1)_\ell, (M - 2)_{\ell + 1}, 1_{\ell + 2}}$ is rather small for realistic values of $N$. For $N \leq 10$, we have the following values for $(N, P_{\mathrm{nontriv}})$: $(3, 0.05)$, $(7, 0.01)$, and $(9, 0.02)$.

All in all, the role of the nontrivial states is either nonexistent [if $M \neq (N + 3) / 2$] or small [if $M = (N + 3) / 2$], and we shall ignore them in the following.

\subsection{Perturbation analysis}
Let us again apply Eq.\ (\ref{eq:schrödinger_n}), this time inside the anharmonicity manifold $\mathcal{A} = \mathrm{span}\{\ket{N_{\ell_N}, M_{\ell_M}} | \ell_N, \ell_M = 1, \ldots, L; \ell_M \neq \ell_N \}$. We can now utilize a lot of the knowledge we have built when analyzing the two previous cases.

\subsubsection{\texorpdfstring{$M = N$}{M = N}}\label{app:two_stacks_interaction_V}
When the stacks are of the same size, $M = N$, the analysis is very much like in the case of only one stack. As long as there are no couplings between the states (at orders $< N$), the boundaries affect in exactly the same way as earlier, but this time separately for each stack. In addition, the proximity of the two stacks also changes the energy of the state.

The zeroth and first orders are trivial. At second order, the state $\ket{N 0 \ldots 0 N 0 \ldots 0}$ is higher in energy than $\ket{0 \ldots 0 N 0 \ldots 0 N 0 \ldots 0}$ due to the presence of the left boundary, and the energy of the state $\ket{N 0 \ldots 0 N}$ is even greater. The energy differences can be calculated simply by adding the single-stack contributions calculated before. Moreover, the energy of the state $\ket{0 \ldots 0 N N 0 \ldots 0}$ is different from that of the state $\ket{0 \ldots 0 N 0 \ldots 0 N 0 \ldots 0}$ as can be seen by direct calculation:
\begin{align}
    \hat{K}_2^{(0)} \ket{N_\ell, N_{\ell + 1}} &= \hbar U \left(\frac{J}{U} \right)^2 2 N \frac{N^2 - 2}{N - 1} \ket{N_\ell, N_{\ell + 1}}, \\
    \hat{K}_2^{(0)} \ket{N_\ell, N_{\ell + m}} &= - \hbar U \left(\frac{J}{U} \right)^2 \frac{4 N}{N - 1} \ket{N_\ell, N_{\ell + m}},
\end{align}
where $m > 1$ and the sites are assumed to be more than one site away from the edges of the chain in order for the boundary effects to not play any role. There is therefore interaction between the quasiparticles, and it is of longer range than the on-site one between the bare bosons. All in all, the anharmonicity manifold $\mathcal{A}$ is split into multiple distinct parts according to the relative positions of the stacks and the boundaries. And a similar thing occurs at every even order below $N$.

Let us study the quasiparticle-quasiparticle interaction at some order $2 n < N$ in more detail. This is manifested by the fact that the state $\ket{N_\ell, N_{\ell + n}}$ has different energy than the states $\ket{N_\ell, N_{\ell + n'}}$ with $n' > n$. The reason for this is again straightforward to understand. Take one boson from each stack and move them first towards each other by $m = 0, \ldots, n$ and $n - m$ sites, respectively, and then back. After the first phase, the two bosons sit at the same site if we start from the state $\ket{N_\ell, N_{\ell + n}}$. But this is not the case for the other states, leading to difference in energy. All the other trajectories generated by $\ham_\mathcal{A}^{(2 n)}$ are the same for both of the states and thus do not contribute to the difference.

To actually calculate this energy difference, we need to consider every possible order we can make the single-boson hops. This is easiest to do by representing the process as a string of length $2 n$ consisting of the letters L and R, where L (R) means that we move the boson originating from the left (right) stack. For fixed $m$, the first $n$ letters should contain $m$ times the letter L and $n - m$ times the letter R, as should also the latter half. For example, the string LRLLLR represent a sixth-order process with $m = 2$, where we first move a boson from the left stack one site to the right, then a boson from the right stack one site to the left, the move the first boson once more to the right and then two times to the left back to the left stack, and finally move the boson from the right stack one site to the right back to where it started from.

First of all, we can write the energy difference $\Delta E^{(2 n)}$ between the states as
\begin{align}
    \Delta E^{(2 n)} &\equiv \braket{N_\ell, N_{\ell + n} | \ham_\mathcal{A}^{(2 n)} | N_\ell, N_{\ell + n}} \notag \\
&\hspace{15pt} - \braket{N_\ell, N_{\ell + n + 1} | \ham_\mathcal{A}^{(2 n)} | N_\ell, N_{\ell + n + 1}} \notag \\
    &= \sum_{m = 0}^{n} \Delta E_m^{(2 n)},
\end{align}
where $\Delta E_m^{(2 n)}$ denotes the energy difference for a fixed value of $m$ defined above. For $m = 0, n$, we obtain quite straightforwardly
\begin{equation}
    \Delta E_0^{(2 n)} = \Delta E_n^{(2 n)} = \hbar U \left(\frac{J}{U} \right)^{2 n} \frac{N^3}{(N - 1)^{2 n - 1}}.
\end{equation}
For $m = 1, \ldots, n - 1$, we need to split the trajectories further. To see this, let us consider strings starting and ending with L. The energy difference now depends on the positions of the first and last occurrences of R in the string. Let $k$ be the index of the first R and $2 n - \tilde{k}$ the index of the last. Furthermore, let $\Delta E_{m k \tilde{k}}^{(2 n)}$ denote the energy difference for such a trajectory. Then, after a while, we obtain
\begin{equation}
    \Delta E_{m k \tilde{k}}^{(2 n)} = - \hbar U \left(\frac{J}{U} \right)^{2 n} \frac{2 N - 1}{2 N - 3} \frac{N^2}{(N - 1)^{2 n - 1}} 2^{k + \tilde{k} - 2 n}.
\end{equation}
We also need to calculate how many trajectories there are with given $m$, $k$, and $\tilde{k}$. A little combinatorics shows that this is given by $\binom{n - k}{m + 1 - k} \binom{n - 1 - \tilde{k}}{m - \tilde{k}}$. The first factor simply counts the number of possible arrangements of the remaining letters $L$ and $R$ in the first half of the chain after fixing the first $k$ letters L$\cdots$LR. And the second binomial coefficient does the same for the latter half of the chain. By performing the analysis also for strings starting and/or ending with $R$, we can write
\begin{align}
    \Delta E_m^{(2 n)} &= - \hbar U \left(\frac{J}{U} \right)^{2 n} \frac{2 N - 1}{2 N - 3} \frac{N^2}{(N - 1)^{2 n - 1}} 2^{- 2 n + 1} \notag \\
    &\hspace{20pt} \times (S_m + S_{n - m})^2
\end{align}
for $m = 1, \ldots, n - 1$, where
\begin{equation}
    S_m = \sum_{k = 1}^{m} \binom{n - 1 - k}{m - k} 2^k.
\end{equation}
Playing around a bit with the geometric series shows that $S_m + S_{n - m} = 2^n$, and so finally
\begin{align}
    \Delta E^{(2 n)} &= \sum_{m = 0}^{n} \Delta E_m^{(2 n)} \\
    &= 2 \hbar U \left(\frac{J}{U} \right)^{2 n} \frac{N^3}{(N - 1)^{2 n - 1}} \left(1 - \frac{n - 1}{N} \frac{2 N - 1}{2 N - 3} \right). \notag 
\end{align}

At $N$th order, we finally obtain a coupling between the states, lifting the remaining relevant degeneracy. Strictly speaking, there is still some degeneracy left. For example, all the states of the form $\ket{0 \cdots 0 N N 0 \cdots 0}$ with $N \geq 3$ which we separated out from the rest of the states at second order have the same energy at sufficiently low orders. The coupling between these is accomplished at order $2 N$, and this would allow the bound pairs to move provided the chain is long enough so that the boundary effects would not intervene. But since the coupling is so weak, the dynamics would be too slow to be observable, and so we ignore the phenomenon here.

All the above can again be condensed into an effective Hamiltonian, which now reads
\begin{equation}
    \ham_{\mathrm{eff}} = \ham_{\tilde{\omega}} + \ham_{\tilde{J}} + \ham_V,
\end{equation}
where the interaction Hamiltonian $\ham_V$ defined in Eq.\ (\ref{eq:quasiparticle_interaction_same_size}) takes into account the interaction between the quasiparticles.

\subsubsection{\texorpdfstring{$M = N - 1$}{M = N - 1}}\label{app:two_stacks_interaction_Xi}
If $M = N - 1$, there is a coupling between the states $\ket{N_\ell, M_{\ell + n}}$ and $\ket{M_\ell, N_{\ell + n}}$ at order $n$. The coupling strength is given by
\begin{align}
    \hbar \, \Xi_n &\equiv \braket{N_\ell, M_{\ell + n} | \ham_\mathcal{A}^{(n)} | M_\ell, N_{\ell + n}} \notag \\
    &= \bra{N_\ell, M_{\ell + n}} \hbar J \hat{a}_\ell^\dagger \hat{a}_{\ell + 1} \notag \\
    &\hspace{60pt}\times \prod_{m = 1}^{n - 1} \hat{W}^{(0)} \hbar J \hat{a}_{\ell + m}^\dagger \hat{a}_{\ell + m + 1} \ket{M_\ell, N_{\ell + n}} \notag \\
    &= (-1)^{n - 1} \frac{N}{(N - 1)^{n - 1}} \hbar \, U \left(\frac{J}{U}\right)^n.
\end{align}
This implies stronger interaction between the quasiparticles than above. 

At first order, for example, the eigenstates of $\ham_\mathcal{A}^{(1)}$ can be divided into two categories. First, we have the linear combinations $(\ket{N_\ell, M_{\ell + 1}} + \ket{M_\ell, N_{\ell + 1}}) / \sqrt{2}$ for $\ell = 1, \ldots, L - 1$ with energies $E^{(1)} = \pm \hbar \Xi_1$. Second, we have the zero-energy Fock states with the stacks farther apart. The superposition states with different $\ell$ couple to each other at order $2 M$, but since this is again greater than $N$, we ignore it here. Similarly, at every order $n < M$ we break the degeneracy between the states where the stacks are $n$ sites apart and the states where the distance between the stacks is greater. Note that this occurs earlier than in the case of equal-size quasiparticles, and so the analysis presented for $M = N$ is not relevant here.

The effective Hamiltonian now reads
\begin{equation}
    \ham_{\mathrm{eff}} = \ham_{\tilde{\omega}}^{(N)} + \ham_{\tilde{J}}^{(N)} + \ham_{\tilde{\omega}}^{(M)} + \ham_{\tilde{J}}^{(M)} + \ham_\Xi,
\end{equation}
where the exchange Hamiltonian $\ham_\Xi$ defined in Eq.\ (\ref{eq:quasiparticle_exchange}) takes into account the exchange coupling discussed above, and we need to include separate single-quasiparticle Hamiltonians for both of the quasiparticles (denoted by the superscripts) since the coefficients $\tilde{\omega}_\ell$ and $\tilde{J}$ depend on the size of the stack.

\subsubsection{\texorpdfstring{$M < N - 1$}{M < N - 1}}
In the case $M < N - 1$, the effective Hamiltonian contains contributions from both $\ham_V$ and $\ham_\Xi$. Following exactly the same reasoning as above in App.\ \ref{app:two_stacks_interaction_V}, we obtain
\begin{align}
    V_n &\equiv \Delta E^{(2 n)} = \braket{N_\ell, M_{\ell + n} | \ham_\mathcal{A}^{(2 n)} | N_\ell, M_{\ell + n}} \notag \\
&\hspace{60pt} - \braket{N_\ell, M_{\ell + n + 1} | \ham_\mathcal{A}^{(2 n)} | N_\ell, M_{\ell + n + 1}} \nonumber \\
&= \frac{N M (N - 1) (N - 3 M + 1)}{(N + M - 3) (N - M) (N - M + 1) (M - 1)^{2 n - 1}} \nonumber \\
    &\hspace{20pt}\times \left(\frac{J}{U} \right)^{2 n} U + \left( M \leftrightarrow N \right),
\end{align}
where the latter term $\left( M \leftrightarrow N \right)$ is just the first term but with $M$ and $N$ exchanged. A general expression for the exchange rate $\Xi_\ell$ is trickier to calculate, but for experimentally relevant values of $N$ and $M$, the neighboring-site exchange rate $\Xi_1$ should be enough. For this, we obtain (cf.\ the analysis in App.\ \ref{app:two_stacks_interaction_Xi} above)
\begin{align}
\Xi_1 &\equiv \braket{N_\ell, M_{\ell + 1} | \ham_\mathcal{A}^{(N - M)} / \hbar | M_\ell, N_{\ell + 1}} \nonumber \\
&= (-1)^{N - M - 1} \binom{N}{M} \frac{(N - M)^2}{(N - M)!} \left(\frac{J}{U}\right)^{N - M} U.
\end{align}

\subsection{Disorder}
\added{
Let us again study the effect of disorder on some dynamical effects. A more detailed analysis can then be made on a case-by-case basis.

First, disorder naturally helps in the formation of bound pairs. If the stacks are of same size, we can simply use the results from the case of a single stack to estimate when the dynamics comes to a halt. The case of different-sized stacks requires a bit more care since we are, in general, unable to make $\hat{P}_\mathcal{A} \ham_D \hat{P}_\mathcal{A}$ to vanish identically simply by adjusting $\omega_\ell$, making the system more sensitive to disorder.

Second, the exchange oscillations within bound pairs respond to disorder like the exchange oscillations in the case of a stack and a single boson. That is, the oscillations start to cease when either $D_\omega$ or $(J/U)^2 D_U$ comes close to $\Xi_n$. To eliminate the disorder in the space spanned by the two states $\ket{N_\ell, M_m}$ and $\ket{M_\ell, N_m}$, we need to set
\begin{align}
\delta \omega_\ell &= \delta U_\ell \frac{N^2 + M^2 + N M - N - M}{2(N + M)} - \delta U_m \frac{N M}{2(N + M)}, \\
\delta \omega_m &= \delta U_m \frac{N^2 + M^2 + N M - N - M}{2(N + M)} - \delta U_\ell \frac{N M}{2(N + M)}.
\end{align}
}

\section{Numerical details} \label{app:numerics}
All numerical results presented in the main text were computed using a numerically exact Krylov subspace method~\cite{saad_analysis_1992} implemented in the Julia programming language~\cite{bezanson2017julia}. The example simulations were computed using the experimentally reasonable parameters $J /2 \pi = \SI{10}{\mega\hertz}$ and $U /2 \pi = \SI{250}{\mega\hertz}$, see e.g. Refs.~\cite{braumuller_probing_2021, ye_propagation_2019}. For the spectrum of Figure~\ref{fig:spectrum}, a larger value of $J / 2 \pi = \SI{20}{\mega\hertz}$ was used for the hopping frequency in order to make the discreteness of the anharmonicity bands more apparent. To elucidate the accuracy of the approximation, in three of the figures of the main text we compare local occupations computed using the full Hamiltonian \eqref{eq:ham} and the relevant effective Hamiltonians.

\section{Disorder tuning example}
\begin{figure}
    \includegraphics[width=\columnwidth]{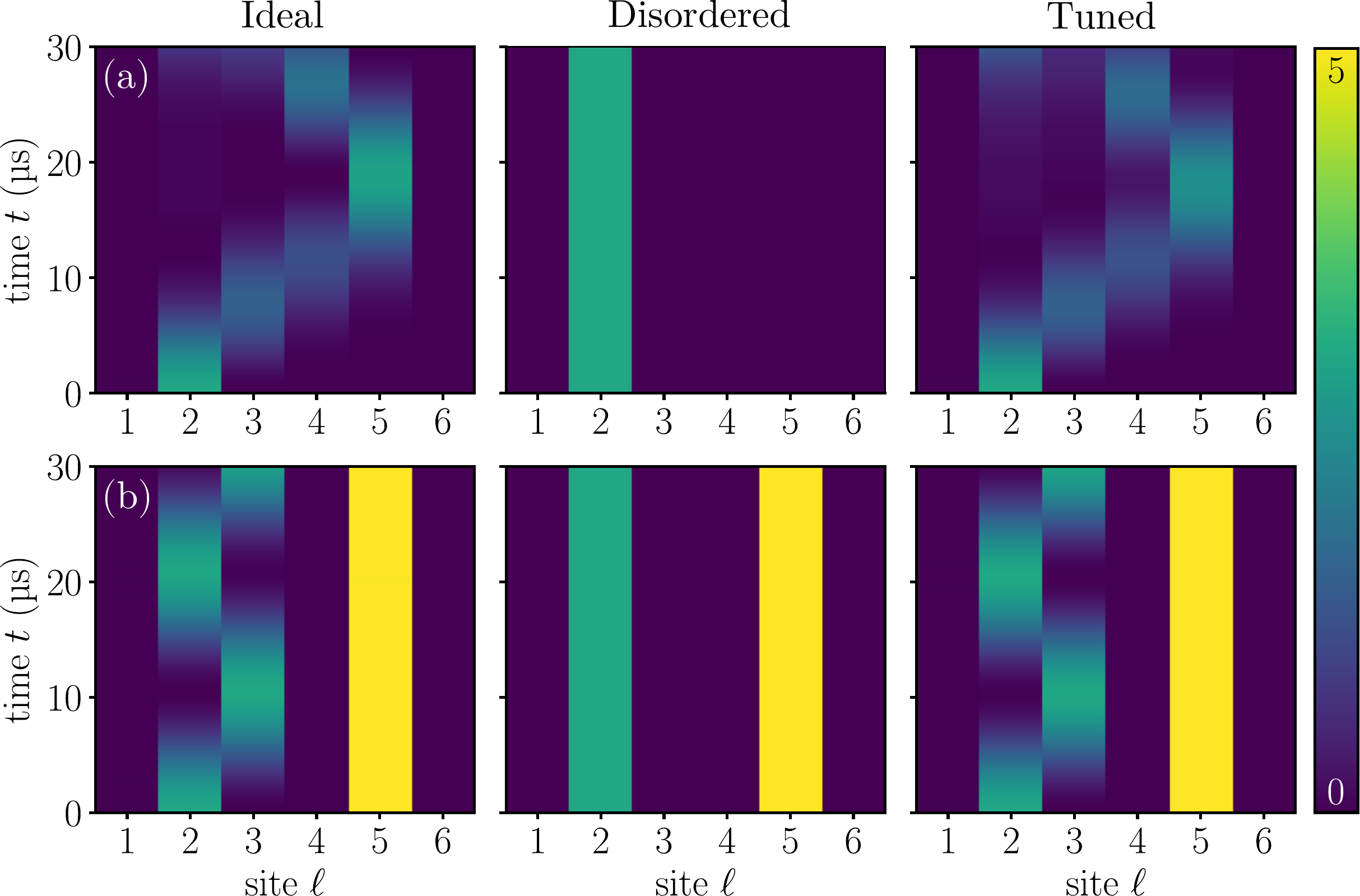}
    \caption{The dynamics frozen by anharmonicity disorder and partial recovery via tuning of on-site energies in a chain of $L = 6$ transmons. (a) A single stack of bosons with the initial state~$\ket{3_2}$. (b)~A~different size stacks with the initial state~$\ket{3_2, 5_5}$. On the left panels, we have the ideal, non-disordered case. The middle panels show the dynamics of a disordered chain. On the right, we have added an on-site energy of $\omega_\ell = -2 \delta U_\ell$, thus approximately restoring the non-disordered dynamics.}
    \label{fig:tuning}
\end{figure}
\added{
As discussed in Sec.~\ref{sec:2D} and in the and in the previous appendices, the effects of anharmonicity disorder can to some extent be remedied via flux tuning. We shall here give a brief simulated demonstration of this in action through two examples shown in Fig.~\ref{fig:tuning}. In the simulations we added a disorder term in Ham.~\eqref{eq:ham}
\begin{equation}
    \hat{H}_D / \hbar = \sum_{\ell = 1}^L \delta U_\ell \hat{n} (\hat{n} - 1),
\end{equation}
with randomly picked $\delta U_\ell \in [-\SI{5}{\mega\hertz}, \SI{5}{\mega\hertz}]$. In the examples, we chose a chain of $L = 6$ transmons, and randomly picked the values $\delta U_\ell/2\pi = \{1.59, -1.75, 4.62, -3.02, 3.81, 3.82\}~\SI{}{\mega\hertz}$. For the sake of simplicity, we assume that any disorder in on-site energies and hopping frequencies are negligible.}

\added{
In the first example we have a single stack of $N = 3$ bosons; see Fig.~\ref{fig:tuning}(a). Clearly, the disorder is strong enough to freeze the dynamics of the stack. We can, however, restore the near-degeneracy of the anharmonicity manifold relevant to a single stack of N bosons by tuning the on-site energy terms, which realizable in superconducting qubits with high-accuracy through magnetic flux controls,}
\begin{equation}
    \hat{H}_{\text{tuning}}/\hbar  = \sum_{\ell = 1}^L \delta \omega_\ell = -\sum_{\ell = 1}^L (N - 1) U_\ell.
    \label{eq:tuning}
\end{equation}
\added{This allows the stack to move again approximately similarly to the ideal, non-disordered case. As discussed before, the dynamics are still modified by the disorder due to the changes to the energies of the hopping paths via the other anharmonicity manifolds. It should be noted that the degree to which this kind of simple tuning restores the ideal dynamics depends on the values of $\delta U_\ell$.}

\added{In the second example, shown in Fig.~\ref{fig:tuning}(b), we have two different size stacks with a stack of five bosons added to the first example. Now, it is not possible to bring the whole anharmonicity manifold to the non-disordered near-degeneracy. Noting however, that the higher stack is, in this example, bound to its initial location by the lower stack, we can restore the non-disordered dynamics by targeting the tuning only on the sites the lower stack can access. We can do this by adding again the tuning term of Eq.~\ref{eq:tuning}.
}
\newpage

\bibliography{main.bib}

\end{document}